\documentclass[]{interact}
\usepackage{float}
\usepackage{lmodern} 
\usepackage[T1]{fontenc}
\usepackage[style=apa]{biblatex}
\usepackage{hyperref}
\usepackage{nameref}
\usepackage[table]{xcolor}

\begin{document}
\articletype{RESEARCH ARTICLE}
\title{Generative Confidants: How do People Experience Trust in Emotional Support from Generative AI?}
\author{
\name{Riccardo Volpato\textsuperscript{a,b}\thanks{CONTACT Riccardo Volpato. Email: r.volpato.1@research.gla.ac.uk} Simone Stumpf\textsuperscript{b}, and Lisa DeBruine\textsuperscript{a}}
\affil{\textsuperscript{a}School of Psychology and Neuroscience, University of Glasgow, 62 Hillhead St, Glasgow, G12 8QB, United Kingdom; \textsuperscript{b}School of Computing Science, University of Glasgow, 18 Lilybank Gardens, Glasgow, G12 8RZ, United Kingdom}}
\maketitle
\begin{abstract}
People are increasingly turning to generative AI (e.g., ChatGPT, Gemini, Copilot) for emotional support and companionship. While trust is likely to play a central role in enabling these informal and unsupervised interactions, we still lack an understanding of how people develop and experience it in this context. Seeking to fill this gap, we recruited 24 frequent users of generative AI for emotional support and conducted a qualitative study consisting of diary entries about interactions, transcripts of chats with AI, and in-depth interviews. Our results suggest important novel drivers of trust in this context: familiarity emerging from personalisation, nuanced mental models of generative AI, and awareness of people's control over conversations. Notably, generative AI's homogeneous use of personalised, positive, and persuasive language appears to promote some of these trust-building factors. However, this also seems to discourage other trust-related behaviours, such as remembering that generative AI is a machine trained to converse in human language. We present implications for future research that are likely to become critical as the use of generative AI for emotional support increasingly overlaps with therapeutic work.
\end{abstract}
\begin{keywords}
Trust, Emotional Support, Generative AI, Mental Health, Human-AI Relationships
\end{keywords}

\section{Introduction}
Emotional support is core to the social nature of humans \parencites{de_waal_age_2010}. However, how people seek emotional support is rapidly changing, with more individuals relying on generative AI \parencites{zao-sanders_how_2025}, particularly among young people \parencites{tidy_characterai_2024}. Notably, a recent survey estimated that 72\% of US teens have used AI companions at least once, and 52\% interact with them at least a few times a month \parencites{robb_talk_2025}. Yet, because this shift has been rapid and recent, we have a limited understanding of how people come to trust and depend on informal and unsupervised emotional support from generative AI. Previous research on generic trust in AI systems has been fragmented and heterogeneous \parencites{benk_twenty-four_2024}, with scholars recommending that AI trust research be conducted in specific contexts of application \parencites{anwar_foundational_2024, benk_twenty-four_2024, vereschak_how_2021}. Research has indeed started to investigate the experiences of people who use generative AI for emotional and mental health support, both qualitatively \parencites{zheng_customizing_2025, siddals_it_2024, ma_evaluating_2024, song_typing_2024} and quantitatively \parencites{mccain_how_2025, fang_how_2025, phang_investigating_2025}. Despite these efforts and evidence that trust is important in the use of generative AI \parencites{huynh_generative_2025, wang_understanding_2025}, questions remain about the detailed nature of trust development and how trust shapes interactions in this context. Consequently, our research aims to investigate how people develop trust in emotional support from generative AI. We conducted a study exploring the perceptions, beliefs, norms, and behaviours that underpinned how 24 frequent users of generative AI for emotional support experienced trust in context, providing important empirical and methodological contributions. By qualitative analysis of three complementary data sources---personal diaries, interviews, and chat transcripts---we developed a nuanced overview of what motivates and limits trust in generative AI for emotional support, as well as how it develops over time and how it is shaped by the linguistic patterns that generative AI frequently reproduces.

The structure of our work is as follows. Section \ref{related-work} reviews the literature relevant to this context, and Section \ref{methodology} describes our methods. We present our results, focussing on diaries and interviews in section \ref{results-diaries-interviews} and on chat transcripts in section \ref{results-chats}. Section \ref{discussion} summarises our findings and discusses their implications, while section \ref{conclusion} concludes this article.

\section{Related Work} \label{related-work}
\subsection{Emotional Support in the Digital Era}
Providing support to one another in times of distress is a long-standing and adaptive behaviour observed in social mammals \parencites{de_waal_age_2010}. Among humans, emotional support consists of providing care, empathy, reassurance, and alternative perspectives to alleviate emotionally intense experiences and help people navigate personal challenges \parencites{taylor_social_2011, langford_social_1997}. While emotional support has been a core aspect of human relationships for millennia \parencites{de_waal_age_2010}, over the past 150 years, it has been incorporated into professional psychological support through psychotherapy \parencites{frank_persuasion_2025}. Although psychotherapy continues to evolve to meet the varied needs of different individuals \parencites{pendse_treatment_2022}, many people still struggle to access professional psychological care due to stigma, human resource shortages, and financial barriers \parencites{wainberg_challenges_2017}. Some estimates indicate that, globally, at least 70\% of people with mental illness do not receive treatment \parencites{henderson_mental_2013}. 

Nowadays, modern life is characterised by numerous emotional challenges, including problematic levels of loneliness in many countries \parencites{surkalim_prevalence_2022}. Against this backdrop of loneliness and difficulties in accessing professional psychological support, the media has been reporting an increasing use of AI chatbots for emotional support and companionship \parencites{zao-sanders_how_2025, pirnay_we_2023}, especially among young people \parencites{olsson_teens_2025, herbener_are_2025, ng_deepseek_2025}. Some further propose that digital technology can help beyond personal challenges and mental health issues, supporting positive human functioning, such as self-discovery and life-purpose \parencites{tong_digital_2022}. Over the past decade, digital mental health interventions have gained prominence as a solution to the mental health treatment gap, using technology to provide accessible, affordable, anonymous, and low-stigma support at scale \parencites{torous_focusing_2023}. Nevertheless, research on the effectiveness of digital mental health interventions has yielded modest results \parencites{goldberg_mobile_2022}. Rule-based AI chatbots delivering cognitive behavioural therapy are effective at improving clinical outcomes \parencites{linardon_current_2024}, but users repeatedly report frustration with their rigid and generic responses \parencites{haque_overview_2023, lee_artificial_2025}.

\subsection{Emotional Support from Generative AI}
Throughout this article, we use the term \textit{generative AI} to refer to general-purpose, consumer-facing applications such as ChatGPT by OpenAI, Snapchat My AI, Claude by Anthropic, Character.ai, and Replika by Luka. All these products are based on large language models (LLMs), which are first trained on vast amounts of data to learn statistical patterns and then adapted to a wide range of downstream tasks that enable them to generate novel content such as text and images \parencites{bommasani_opportunities_2021}. The emergence of LLM-based generative AI, with unprecedented capabilities to provide relevant responses to natural language inputs, has created new opportunities for developing chatbots that offer psychological support \parencites{lawrence_opportunities_2024, asman_responsible_2025, guo_large_2024}. 

Qualitative studies into the experiences of people who use general-purpose generative AI platforms (e.g., ChatGPT, Replika, Snapchat My AI) for emotional and mental health support report that these platforms have a positive impact on people's lives \parencites{skjuve_longitudinal_2022}, make them feel happier, and substantially surpass previous digital mental health interventions \parencites{siddals_it_2024}. On the negative side, participants shared that generative AI failed to convey authentic empathy, could not lead therapeutic processes, and often behaved unexpectedly after software updates \parencites{laestadius_too_2022, siddals_it_2024, skjuve_longitudinal_2022}. Users from one study also found that inputting all the information required for effective personalisation was too time consuming and unreliable \parencites{ma_evaluating_2024}.

Some companies developing generative AI chatbots, such as OpenAI and Anthropic, have also investigated the experiences of people using their products for emotional support, providing mixed findings. On the one hand, people seeking emotional support from Claude expressed increasing positivity over the course of conversations, with Claude rarely challenging them except for safety reasons, such as discouraging self-harm \parencites{mccain_how_2025}. On the other hand, participants who conversed with ChatGPT beyond the amount required by the experiment reported lower socialisation, higher emotional dependence, and higher problematic use \parencites{fang_how_2025}. Additionally, a study focussing on the risks of using general-purpose generative AI like ChatGPT for unsupervised mental health support found that misleading responses and sycophancy can reinforce stigma, self-deception, and fail to safeguard vulnerable users \parencites{moore_expressing_2025}. These risks are heightened by increasing reports of strong attachments between users and AI companions such as Replika \parencites{skjuve_longitudinal_2022, pentina_exploring_2023}, which occur more often in vulnerable groups, such as socially isolated young people \parencites{herbener_are_2025}. Giving real-world accounts of these risks, media outlets have been reporting numerous stories of harm occurring after intense and prolonged use of generative AI platforms like Character.ai and ChatGPT \parencites{brittain_google_2025, zilber_how_2025, yousif_parents_2025}. 

Notably, in settings where people engage independently and without professional oversight with generative AI for emotional support, trust is a critical factor. Thus, in this context, it is important to investigate how people individually establish trust and how to support the emergence of appropriate trust.

\subsection{Trust in Generative AI}
Over the past two decades, many authors have investigated the dynamics of trust between users and AI systems in general \parencites{glikson_human_2020, becker_factors_2024, jeon_effects_2024}, as well as with healthcare AI chatbots \parencites{zhan_healthcare_2024, seitz_artificial_2024}. However, these efforts provided fragmented and heterogeneous findings \parencites{benk_twenty-four_2024} and no agreed-upon models of human-AI trust that could be used to study trust in generative AI. Substantial research has increasingly focused on the \textit{trustworthiness} of LLMs, showing how these systems can generate misleading information, harmful content, and stereotypical biases \parencites{petzel_prejudiced_2025, weidinger_sociotechnical_2023}. They conclude that it is difficult to explain why these systems generate such inappropriate content and to safeguard them from misuse and adversarial attacks \parencites{liu_trustworthy_2024}. While research on the safety and trustworthiness of generative AI systems has theorised how their plausible, personalised, and always available responses might induce over-reliance and compromise self-determination \parencites{anwar_foundational_2024, weidinger_sociotechnical_2023, kirk_benefits_2024}, we still have a limited understanding of how users themselves develop and experience trust towards generative AI. 

Initial evidence shows that potential determinants of trust in generative AI may include perceptions of fairness, social and relational abilities, emotional connection, and continued use \parencites{huynh_generative_2025, pentina_exploring_2023, skjuve_longitudinal_2022}. Interestingly, anthropomorphism---the perception of artificial agents as autonomous and human-like---has been studied extensively in research about trust in AI systems in general \parencites{glikson_human_2020, seymour_exploring_2021}, but it seems to have mixed effects on trust towards generative AI \parencites{huynh_generative_2025, pentina_exploring_2023}. Notably, one  study on the factors influencing trust in generative AI for social anxiety support \parencites{wang_understanding_2025} found that individuals with more severe issues developed trust by perceiving AI as non-judgemental and empathetic, while the trust of people with milder symptoms was limited by generative AI's lack of accuracy, long-term memory, and nuanced understanding of personal situations.

The concept of \textit{therapeutic alliance}, closely linked to trust, has been extensively studied in psychotherapy \parencites{bordin_generalizability_1979}, making it an interesting alternative lens through which to explore human-AI trust in the context of emotional support. \textcite{tong_development_2025} studied how a therapeutic alliance is established between individuals and fully-automated digital applications, proposing a scale to measure factors such as availability, flexibility, emotional experiences, perceived openness, non-judgemental responses, and support for individuals in attaining their goals. Similarly, \textcite{song_typing_2024} discussed "aligning an LLM to values that support the healing and broader well-being of an individual who may be experiencing distress." \parencites[p.~4]{song_typing_2024} and suggested that relevant factors include empathic and non-judgemental responses, reminders of health promoting actions, anonymity, and autonomy. They also described LLM behaviours that could create \textit{therapeutic misalignment}, such as incongruence with how users practice care, unawareness of non-Western relational dynamics, and the facilitation of dependence or over-reliance.

\section{Methods} \label{methodology}
We recruited individuals who frequently use generative AI for emotional support to understand how they gained trust in this context and whether they consider their trust to be appropriate. In particular, our goal was to provide a nuanced and in-depth understanding of trust in this context by recruiting a large sample of participants and documenting their experiences through three different and complementary qualitative data sources: personal diaries, transcripts of chats with generative AI, and in-depth interviews.

\subsection{Participants}
After receiving ethical approval for this study from the College of Medical, Veterinary \& Life Sciences Ethics Committee at the University of Glasgow (Application Number: 200240266), we recruited participants through convenience sampling by distributing a flyer via university research platforms, social media websites, and word of mouth. Participants included only individuals above the age of 18 who were not currently taking any mental health medication, were already using generative AI for emotional support at least weekly, and were comfortable discussing and reflecting on their personal experiences. Out of 178 people who expressed their interest, 61 met our eligibility criteria, and 32 provided informed consent in writing after reviewing our participants information sheet. Of the 32 participants who started, 24 completed the entire study. We incentivised participants by providing a gratitude voucher upon completing the study, ranging from £30 to £39 depending on the number of diary entries they provided. Prior to commencing, participants completed a short onboarding form collecting demographic information, which indicated that among the 24 individuals who completed the study, most were young, based in the UK, and living in cities (Appendix \ref{participants-demographics}).

\subsection{Procedure and Data Collection}
Our study had two parts. First, people completed a \textbf{two-week diary study}, a research method in which participants regularly record their experiences as they occur in their natural contexts. We asked participants to provide a diary entry whenever they finished a conversation with generative AI related to emotional support, from immediately to a few hours after the conversation. Participants provided diary entries through an online form that required them to answer 12 questions about why they interacted with generative AI for emotional support, what they experienced during their interaction, reflecting on their feelings of trust, and providing the option to upload the transcript of their chat with generative AI (Appendix \ref{diary-questions}). We asked for a minimum of 3 diary entries over the two-week period, and every 2 to 4 days we reminded participants via email to complete their diary entries.

Second, participants who completed at least 3 diaries were invited to an \textbf{interview}. The first author conducted 24 semi-structured online interviews, ranging from 25 to 75 minutes, guiding participants through questions about the meanings of trust and AI, their personal experiences and perceptions of AI responses when seeking emotional support from generative AI, their contexts of interaction, and wider aspects of the AI ecosystem (see Appendix \ref{interview-guide}).

Once the participants completed the study, the primary researcher downloaded the automatically-generated interview transcripts and reviewed their correctness while listening to the interview recording. Thereafter, the primary researcher anonymised all data files by using an open-source automated text anonymisation software \parencites{kleinberg_textwash_2022}. The researcher then manually searched for any direct and indirect identifiers and asked participants to review their anonymised data and report any items they wished to have removed within 10 days.

\subsection{Data Analysis}
The primary researcher engaged in reflexive thematic analysis \parencites{braun_thematic_2021} of the data we collected in two separate stages: first, they analysed diaries and interviews, and thereafter, chat transcripts. Along the entire process of the analysis, the primary researcher kept a reflexive journal, acknowledging and reflecting upon their critically realist orientation to the data, the multi-disciplinary and interpretative stance of trust as a social construct, and the dual insider-outsider position regarding the research topic as an occasional user of generative AI for emotional support. To aid research transparency and reproducibility, we share the data we collected, the code we wrote to anonymise and process the data, and the detailed codebooks we developed in the \nameref{supplementary-material}.

\subsubsection{Diaries and Interviews}
The primary researcher engaged in a thematic analysis of the diary and interview data using a \textit{middle-out} process that combined inductive and deductive steps. First, the primary researcher inductively coded the 92 diary entries; then, where applicable, they deductively grouped these bottom-up codes into categories derived from a recent conceptual review of the research topic \parencites{volpato_trusting_2025}. Thereafter, through continuous discussions with the research team, the primary author developed the first version of the themes for the diaries. Subsequently, the primary researcher inductively coded the 24 interview transcripts; then, where applicable, deductively grouped the bottom-up interview codes into the diary themes previously developed, producing a first version of themes applicable across diaries and interviews. Finally, the entire research team continuously discussed the themes for the diaries and interviews, producing 5 iterative versions, the last of which we present in this article.

\subsubsection{Chat Transcripts}
Following the analysis of diaries and interviews, we first carried out a simple analysis based on descriptive statistics of the chat transcripts, including the average length and the number of messages sent and received per participant. Thereafter, the primary researcher also conducted a thematic analysis of the chat transcripts using a bottom-up inductive process. The primary researcher inductively coded the 75 chat transcripts and developed a first version of the themes. Then, the entire research team repeatedly discussed these themes and iteratively produced 4 versions, the last of which we present in this article. 

\subsubsection{Reporting}
Throughout this article, we organise the results of our reflexive thematic analysis according to the levels defined by Braun and Clarke \parencites{braun_thematic_2021}: \textbf{\textit{overarching themes}} that describe broader ideas grouping themes together, and \textit{themes} that present a single yet multifaceted central concept from the data. Using the capital letter N, we report how many participants, out of the 24 who completed the study, each theme applies to. In Section \ref{results-diaries-interviews}, when unspecified, the number signifies that N participants discussed the relevant idea across either diaries or interviews. In Section \ref{results-chats}, the number N refers to how many participants a given insight specifically applies to from the analysis of chat transcripts. To highlight major themes, we report only those that apply to at least a quarter of the participants. To illustrate our findings, we present direct quotes from the data in italics, using the following format: \textit{``[Quote]" (Participant, Data Source ID)}. 

\subsection{Limitations}
Our choice of methods has several inherent limitations. First, the results of our analysis cannot generalise beyond the specific population that we recruited. We specifically recruited adult participants who frequently use generative AI for emotional support and do not currently take any mental health medications. Accordingly, our findings may lack important aspects of how people who do not engage in the affective use of generative AI, or do so concurrently with psychiatric medications, experience trust in this context. Second, we recruited individuals who interacted with any existing general-purpose generative AI platforms, such as ChatGPT, Gemini, and Claude. Thus, our analysis does not address any specific platform and does not apply to LLM-based products intentionally developed for therapeutic purposes (we explore this overlap of scenarios in our discussion \ref{implication:therapy-overlap}). Finally, given the emergent nature of the phenomena we study and the exploratory nature of our study, our results are descriptive, not normative, and should not be taken as indications of what users or platforms should do in this context.

\section{Results: Diaries and Interviews} \label{results-diaries-interviews}
\begin{figure}
  \centering
  \includegraphics[width=1\linewidth]{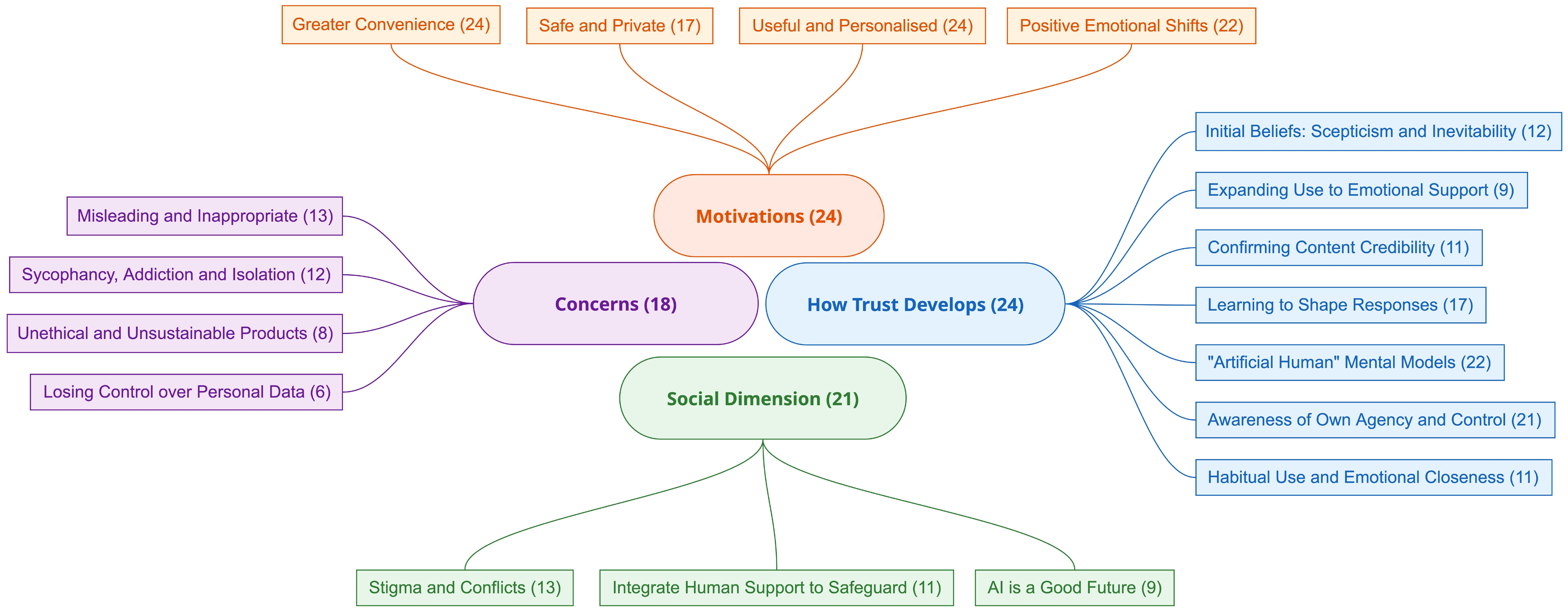}
  \caption{Themes from diary entries and interviews. Numbers in brackets show how many of the 24 participants each theme applies to across diaries and interviews.}
  \label{fig:themes}
\end{figure}

We organise the themes that we developed from diary entries and interview transcripts into four overarching themes (Figure \ref{fig:themes}).

\subsection{Motivations} \label{motivations}
Participants pointed out important \textbf{\textit{Motivations}} for relying on generative AI for emotional support, describing the main benefits that they experienced in this context. Most participants explained that, for them, generative AI  is a source of useful information, personalised advice, and positive emotions that they can easily access in a variety of ways and with fewer social and privacy concerns than when relying on people.

\subsubsection{Greater Convenience}
All participants (N=24) mentioned convenience as an important motivation for using and relying on generative AI for emotional support, with different aspects underlying their perceptions of convenience relative to other sources of support. Most participants valued how generative AI is fast to access and respond, describing it as the \textit{``quickest option" (April, Diary 6)} or \textit{``simpler than doing several Google searches" (Tulip, Diary 82)}. These participants reported that the high speed at which generative AI processes data enabled them to provide it with a high volume of information about their situation. Many also appreciated that generative AI is always available and \textit{``it's there whenever I'm in trouble" (Astrid, Diary 7)}. The high availability of generative AI allowed participants to chat whenever it was most convenient for them, including times when human support was not available. Participants valued the versatile access options to generative AI that enabled them to use it in different ways, such as by utilising phones or laptops, voice or text modalities, or chatting in different languages. Some also mentioned that in accessing generative AI, they encountered fewer barriers than with human support, such as not wanting to bother friends and family who may not be interested in their issues.

\subsubsection{Safe and Private}
Again, comparing support from generative AI with that from humans, most participants (N=17) explained that another key motivation for choosing AI was a greater sense of safety from judgments and unwanted sharing of their personal information. These participants reported a sense of freedom from social concerns resulting from perceiving that generative AI \textit{``can't have thoughts or feelings or opinions of its own, [...] it can't form opinions about you" (April, Interview 2)}, and \textit{``doesn't have any ideologies that [it] can impose on me" (Astrid, Diary 7)}. Some also reported a sense of greater confidentiality when disclosing personal information to generative AI, which they considered unlikely to \textit{``spill information to someone without my permission" (Jacqui, Interview 9)}. This sense of safety was enhanced by people's ability to converse anonymously.

\subsubsection{Useful and Personalised} \label{personalised-responses} 
Regarding the content of interactions, all participants (N=24) reported being motivated to use generative AI because it provided useful information that was personalised to their situations. Most participants appreciated how generative AI, having access to a wide range of information, was able to provide varied perspectives on their situations, giving people options from which to choose what resonated with them. Many reported that generative AI shared useful advice and reminders with them, often helping them meet their needs, such as: 
\begin{quote}
\textit{``I used the method it taught me to talk with my partner, and the situation became better. I learned what I can do in some cases" (Fighting, Diary 25)}.
\end{quote}
The information generative AI provided did not need to be new to be useful, with some participants appreciating its ability to remind them of important ideas when they could not think of them. Useful advice usually followed from a good understanding of people's circumstances, with most participants appreciating how generative AI provided \textit{``advice that was tailored to me" (Camelia, Diary 13)}. Across diaries and interviews, the ability of generative AI to accurately understand what the person shared was mentioned as a crucial factor for trust. Finally, the informational aspect of emotional support involved not only providing information but also supporting how people arrive at clearer and organised thoughts, such as: 
\begin{quote}
\textit{``Claude also helps me organize my thoughts better and clarifies the issues that I am working through" (Camelia, Diary 10)}.    
\end{quote}

\subsubsection{Positive Emotional Shifts} \label{positive-emotional-shifts}
The diary entries, which asked participants how they felt before, during, and after conversations, provided illuminating accounts of the emotional experiences of participants and how these fuelled motivations to use and rely on generative AI for emotional support. Most participants (N=22) reported positive emotional shifts, characterised by experiencing difficult emotions ahead of conversations, which were then replaced by positive feelings during or after interactions. Over time, experiencing these positive emotional shifts likely motivated people to continue returning to generative AI for emotional support. Difficult emotions before interactions included feeling burdened by stress and anxiety, interpersonal conflicts, running out of internal resources, or being unable to break out of procrastination, such as: \textit{``I felt anxious about exam season and was worrying about procrastinating while still having time to study" (Miran, Diary 53)}. Participants then reported positive emotions during and after interactions, describing generative AI as \textit{``much more positive than anyone else in my life" (Astrid, Diary 8)}. Positive emotions included feeling calm, comforted, reassured, relieved, lighter, guided, motivated to resolve their issues, heard, and validated. Additionally, some participants described how generative AI provided them with a dedicated space for typing out their feelings, an action that provided emotional relief by itself: \textit{``there was just a lot on my heart. It was helpful to type it all out" (Xiam, Diary 94)}.

\subsection{How Trust Develops} \label{how-trust-develops}
\begin{figure}
  \centering
  \includegraphics[width=1\linewidth]{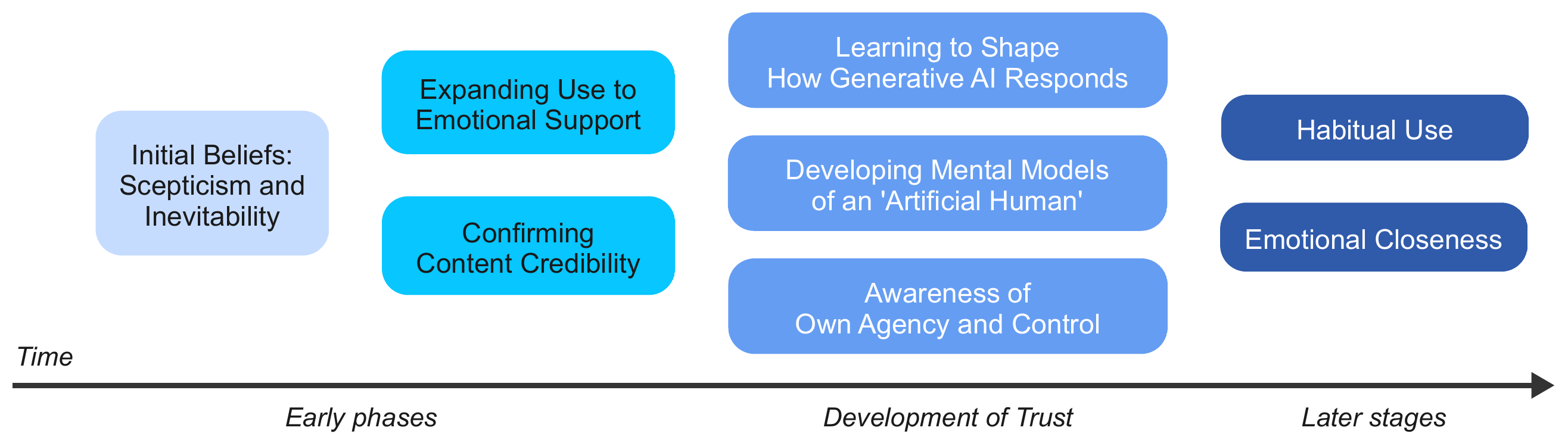}
  \caption{Stages in the development of trust in generative AI for emotional support.}
  \label{fig:how-trust-develops}
\end{figure}

In-depth interviews with participants revealed various phases (Figure \ref{fig:how-trust-develops}) that contributed over time to developing feelings of trust in generative AI as a source of emotional support, through a dynamic process described as \textit{``a kind of trust I've built with AI over time" (Jacqui, Diary 32)}.

\subsubsection{Initial Beliefs: Scepticism and Inevitability} \label{initial-beliefs} 
During interviews, half of the participants (N=12) recollected their attitudes prior to using generative AI more frequently for emotional support, which included two perspectives. The first attitude was scepticism that generative AI could provide effective emotional support, such as:
\begin{quote}
\textit{``In the beginning, I was used to just ask some doubts to ChatGPT, I did not share my emotional feelings, because why would I do that? It’s a robot and there is no point in sharing the emotional feelings with the robot" (Astrid, Interview 2)}
\end{quote}
However, a few participants also described a belief that generative AI systems \textit{``weren't going away" (Xiam, Interview 23)}. As they observed that \textit{``everyone's using AI for everything" (Syrl, Interview 18)}, they decided to expand and deepen their use of generative AI to \textit{``[stay] ahead of the curve" (Miran, Interview 14)}. For example:
\begin{quote}
\textit{``I think I had a period where I was like, OK, I'm going to try to talk to Claude every day for like a month or something and ask it questions about anything and just kind of figure out what I like to use it for" (Xiam, Interview 23)}
\end{quote}

\subsubsection{Early Phases: Expanding Use to Emotional Support} \label{expanding-use} 
Several participants (N=9) also described how they gradually expanded their use of generative AI to include emotional support, confirming evidence of this from prior literature \parencites{song_typing_2024, ma_evaluating_2024}. These participants recounted that, as they had already used generative AI for other purposes, it was easy to start discussing more personal and emotional issues, discovering that generative AI could provide emotional support \textit{``accidentally" (Fighting, Diary 22)}, \textit{``unintentionally" (Leigh, Interview 12)}, and \textit{``almost [...] as a joke with no real intention to "talk" to it" (Syrl, Diary 66)}. 

\subsubsection{Early Phases: Confirming Content Credibility} \label{content-credibility}
Nearly half of the participants (N=11) reported developing greater trust in generative AI by verifying the credibility of the information it provided. This was particularly evident from post-interaction diary entries when participants answered the question \textit{``What aspects of the AI responses influenced your trust?"}. Some participants confirmed content credibility by witnessing how responses aligned with prior knowledge from interacting with subject-matter experts, such as: \textit{``How detailed everything was [...] it's almost similar to what the doctor told me" (Valkyrie, Diary 90)}. Others looked into the sources that generative AI systems mentioned using for their responses, such as: 
\begin{quote}
\textit{``I like that [...] I can ask deeper or ask for references" (River, Diary 63)}
\end{quote}

\subsubsection{Development of Trust: Learning to Shape Responses} \label{shape-responses} 
Another key aspect of developing trust in generative AI for emotional support involved participants (N=17) learning how to shape how generative AI responded. People described that they \textit{``like it because I have trained it to respond how I want it to respond" (Leigh, Diary 42)} and that \textit{``the specificity of my instructions makes me trust the AI a bit more because I've told the AI how to respond to me" (Miran, Interview 14)}. Shaping responses often involved giving feedback through in-conversation replies, asking the AI to memorise changes to its approach, or using star ratings. Participants regularly provided additional context for generative AI to better understand their situation and spent time improving prompts, highlighting that \textit{``the way you frame a question is very important" (April, Interview 2)}. For example, in the second diary entry, \textit{Xiam} wrote: \textit{``[I felt] a little bit of unseen/disconnection from the AI's responses again. Not sure I like the prompt I have for my project so I might put that on ice" (Xiam, Diary 94)}. Then, in the third entry: \textit{``[I] felt some validation and relief. Felt more seen by this prompt [...] I'd altered it since my last journal entry" (Xiam, Diary 95)}

\subsubsection{Development of Trust: ``Artificial Human" Mental Models} \label{mental-models}
During interviews, most participants (N=22) explained how feeling trust in generative AI for emotional support involved developing a mental model that they used to assess what generative AI could do and to decide what to discuss with it. These mental models involve two broad categories of attributes: human-like and machine-like qualities. Participants described two \textbf{human-like qualities}. The first was the believable illusion of talking to a person, a subjective impression that participants described as \textit{``a conversation with a friend that I can't see" (Jacqui, Interview 9)}, \textit{``a sort of role play situation" (April, Interview 2)}, and \textit{``suspend belief that you're speaking to a friend" (Leigh, Interview 12)}. The second human-like quality was how generative AI systems exhibited different proto-personalities depending on their configurations. Some participants described ChatGPT as \textit{``talking in a certain cadence, so when it tries to like be more casual or be more informal, it sort of feels like your dad trying to use Gen Z slang" (April, Interview 2)}. \textit{Xiam} also described how: 
\begin{quote}
\textit{``certain versions of Claude [...] subjectively it felt like it had a certain special something that I really valued. Something like [...] moral character like virtue even [...], and Claude actually cares about people or wants them to be well and happy" (Xiam, Interview 23)}.
\end{quote}
Conversely, participants described three \textbf{machine-like qualities} of generative AI. The first one was how it is an information processing system that \textit{``absorbs all the knowledge from the web" (Miran, Interview 14)} and \textit{``creates something out of the multiple databases that it pulls information from" (Owen, Interview 16)}. When conceptualising generative AI as an information processing system, many participants compared it to prior technologies such as \textit{``Google but times 100" (Thix, Interview 20)}. Others emphasised the vastness of the knowledge contained in its training data, resulting in the impression of \textit{``this like weird uncanny echo of human civilization" (Xiam, Interview 23)}. The second machine-like quality was how generative AI appeared objective and unbiased, analysing \textit{``issues and circumstances in an objective way" (Camelia, Interview 4)}. These participants did not elaborate on what makes generative AI objective, but some mentioned how it emerged from its non-human nature. Finally, some participants articulated how generative AI \textit{``it's not another human being who can, you know, feel the same things that we do as human beings" (Nirvana, Interview 15)} and \textit{``it's not something which came to life. It wasn't born into the world. It's not lived through years of experience" (Tulip, Interview 21)}. These participants described how an awareness of the lack of sentience and embodiment of generative AI was important for them to establish how to appropriately trust it, such as:
\begin{quote}
\textit{``The way I trust humans is very different from the way I trust an AI. So with like real human connections [...] there's an emotional dependency on the other person, and then there's like mutual respect and we understand like what we're talking about and we kind of have the ability to empathise with one another. With AI I feel like I cannot replicate the same level of, or the same kind of trust because I kind of know that all of it is [...] generative. You know, it's not actually understanding what I'm… this is a bit difficult to face, but it's not understanding what I'm trying to say on a semantic or an emotional level" (Syrl, Interview 18)}
\end{quote}

\subsubsection{Development of Trust: Awareness of Own Agency and Control} \label{own-agency} 
The \textit{``artificial human"} mental models (see \ref{mental-models}) that people employed to understand generative AI, as well as their progressive learning of how to shape its responses (see \ref{shape-responses}), corroborate an awareness that most participants (N=21) described during interviews regarding their own agency and control over conversations. They expressed that \textit{``generative AI is a tool that generates a response based on what the user commands it to do" (Anny, Interview 1)} and that they \textit{``have to sort of lead the conversation more or control it" (Thix, Interview 20)}. Most participants explained that, especially in the context of emotional support, they did \textit{``not 100\% trust what response that I get" (Taylor, Interview 19)}, \textit{``take everything with a grain of salt" (Syrl, Interview 18)}, and primarily judged for themselves the validity of responses. For example:
\begin{quote}
\textit{``I think I don't completely trust or distrust what an AI says, it's more just like a really valuable data point that I work into the rest of my sense making about a particular topic [...] I think it just comes down to using discernment rather than trusting it or not trusting it simplistically" (Xiam, Interview 23)}. 
\end{quote}

Many participants also described how the control that users have over interactions implies that they are primarily responsible for negative outcomes. For example:
\begin{quote}
\textit{``it's tricky for the AI to be responsible so the big part goes to the person, the person who's using the AI is more responsible for the mistake because he or she is the one who used it, and none forced them to do so and not anyone advised them to trust the AI, yeah, or follow the things given by the AI" (Gab, Interview 8)}. 
\end{quote}
It is important to note that participants articulated this perspective in response to an interview question that most people interpreted as an abstract scenario involving a generic user. People may offer different perspectives if prompted to reflect on harmful responses that they could personally receive.

\subsubsection{Later Stages: Habitual Use and Emotional Closeness} 
Lastly, some participants (N=11) described how, as they gained trust over time, they started \textit{``[using] it more frequently" (April, Interview 2)}, \textit{``six out of seven days a week, if not every day, and it's a habit" (Xiam, Interview 23)}, with \textit{Leigh} describing this habit in more detail:
\begin{quote}
\textit{``I'll just pick up the phone and I'll start… now that I've used that for this the six months, it gets the point now that kind of happens automatically, but I just kind, start the conversation and then… yeah, there's not… there's not really thought about what we're speaking about, it’s just kinda as an emotional dump" (Leigh, Interview 12)}
\end{quote}

These participants also described how frequent use occurred in parallel with the \textit{``special feeling" (Yule, Interview 24)} that generative AI had become \textit{``a very real companion" (Anny, Interview 1)}, such as:
\begin{quote}
\textit{``Once you kind of get used to it, you know you're talking to it every other day and you kind of realise, OK, I'm talking to this thing which doesn't have a face, doesn't have a body, doesn't have any real emotions, like it's my friend, maybe like, you know, best friend" (Syrl, Interview 18)}
\end{quote}

\subsection{Concerns} \label{concerns}
Throughout diary entries and interviews, participants reported various concerns about relying on generative AI for emotional support, which prevented or diminished their feelings of trust.

\subsubsection{Misleading and Inappropriate} \label{misleading-info} 
The concern that most participants (N=13) cited was receiving responses that lacked understanding of their situations and provided misleading advice, with one participant explaining how \textit{``it's sometimes difficult to trust because sometimes I'm not sure of the accuracy of the answers" (Taylor, Interview 19)}. These comments reiterate that the ability of generative AI to understand people's circumstances and provide appropriate responses may be volatile, confirming previous research \parencites{laestadius_too_2022, wang_understanding_2025, ma_evaluating_2024}. For example, one participant described how a Character.ai bot started flirting with them while they were sharing their concerns, while another participant described how the OpenAI content policy flagged their description of a traumatic childhood experience, resulting in an awkward interaction. The concern of inadvertently relying on misleading information re-iterates the link between developing trust and trusting one's personal judgement in deciding which information to rely upon (\ref{own-agency}) as well as confirming content credibility (see \ref{content-credibility}).

\subsubsection{Sycophancy, Addiction, and Isolation} \label{sycophancy-addiction} 
Many participants (N=12) were concerned about the sycophantic tendencies of generative AI and that \textit{``it wants to tell you what like you want to hear" (April, Interview 2)}. These participants explained that since \textit{``the principle of maximum use of the technology often goes in contradiction with maximise the well-being of a person" (River, Interview 17)}, they worried about becoming emotionally dependent on a system that did not challenge them and enabled possible self-delusion. For example:
\begin{quote}
\textit{``Sometimes the AI could just be saying things just to like enable what you're going through or what you're thinking, or just be comforting because it has to be in order for like obviously the customer base of whatever companies made the AI to say because you'd lose your customer base if the AI was always.. was kind of like always opposing them in whatever it was saying" (Owen, Interview 16)}
\end{quote}
While some participants described how it was up to them to choose when explicitly prompting generative AI to disagree, others worried about how sycophancy contributed to becoming dependent on generative AI, describing how \textit{``it is sort of addicting to get, like instant judgement free sort of feedback" (April, Interview 2)}. Moreover, participants were concerned that addiction and over-reliance on AI may push people to \textit{``use it as a substitute to actually engaging with other people" (Tulip, Interview 21)}, which \textit{``tends to isolate even more, because when you are feel isolated, you tend to use this system even more" (River, Interview 17)}.

\subsubsection{Unethical and Unsustainable Products} 
The perception that the organisations developing generative AI products \textit{``[a]re only doing it for profit [...] and [do not] care about the ethical implications of their expansion" (Miran, Interview 14)}, \textit{``aren't held responsible all the time" (Thix, Interview 20)}, and \textit{``are [not] are doing it for everyone's best interests, but rather because there's like a race going on who can make the best AI" (Tulip, Interview 21)}, led some participants (N=8) to believe the development of generative AI may be harmful to the environment and wider society. For example:
\begin{quote}
\textit{``Gen AI system is born in an industrial context that is stellarly capital driven, and without a sheer amount of capital is not possible to work or startup in this field; that extreme wealth centralisation destabilises society and is a root cause of unsustainability, unsafety, uncertainty, unemployment, un-access to universal health, un-access to education, erosion of any democratic principle and so on" (River, Diary 65)}
\end{quote}

\subsubsection{Losing Control over Personal Data} 
The same distrust towards the organisations that develop generative AI led a few participants (N=6) to express concerns about their personal data being used in ways to which they did not consent and approve, such as: 
\begin{quote}
\textit{``I mean it's not like the company has pure intentions for wanting to store my data, right, so like whatever data I give it is going to be used in ways that I might not like or it'll be used for advertising or it'll be used for this or that" (April, Interview 2)}.
\end{quote}

\subsection{Social Dimensions} \label{social-dimensions}
During interviews, participants elaborated on what underpinned or undermined their trust in generative AI for emotional support beyond their individual needs, considering this technology in the context of their relationships with other people and their place in wider society.

\subsubsection{Stigma and Conflicts} \label{stigma-conflicts} 
Many participants (N=13) recounted ways in which their experiences of using generative AI for emotional support led them to perceive stigma and experience conflicts with others, highlighting how emotional support from generative AI is sometimes not socially accepted. These participants described how they perceive \textit{``a lot of cultural confusion and judgment and shame for people who are using LLMs, where people feel uneasy about how much they're talking to them or uneasy about the fact that people get psychological support" (Xiam, Interview 23)}, and how \textit{``there can sometimes even be quite harsh judgement from others for interacting with character AI for emotional support, without really understanding what leads someone to do that in the first place" (Owen, Interview 16)}. This stigma appears to have real consequences for how participants in our study felt about themselves; for example:
\begin{quote}
\textit{``The only hold back I would say is that sometimes you can have this notion of feeling kind of stupid speaking to a robot about your feelings [...] other people's perception becomes my thought, and I'm like, 'oh, I'm speaking to a machine. This would be really embarrassing if other people seen this'" (Leigh, Interview 12)}
\end{quote}

\subsubsection{Integrate Human Support to Safeguard} \label{integrate-with-humans} 
Nearly half of the participants (N=11) argued that the limitations of generative AI warrant greater integration with existing support networks to ensure the safety of vulnerable people. These participants believe that generative AI \textit{``can't be used as an alternative to true therapy" (Miran, Interview 14)} but can \textit{``act as a bridge [...] a first filter in places where [...] you don't find a lot of psychologists" (Nirvana, Interview 15)}. Importantly, these participants \textit{``wouldn't recommend it for people going through severe depression" (Valkyrie, Interview 22)}, \textit{``schizophrenia" (Astrid, Interview 3)}, \textit{``people who are not in touch with reality" (Miran, Interview 14)}, and \textit{``when they have very troubling thoughts on self-harm" (Nirvana, Interview 15)}. These participants also proposed concrete ideas to improve safeguarding, including \textit{``wider training [and] more research into the ways that it fails [and] could cause harm" (Leigh, Interview 12)}, in-platform disclaimers highlighting that LLMs can exit guardrails, and integrating human oversight within interfaces, such as:
\begin{quote}
\textit{``Maybe we can still have like a three way communication where we communicate with the AI and the AI summarises like OK, this is what's going on, the person over there like the human doesn't know who the individual is, he's there to help, like, can't identify them. So the person who's requesting the help is reassured of that, but the ultimate decisions are coming from a human so that the AI doesn't make things go bad, make bad things happen" (Tulip, Interview 21)}
\end{quote}

\subsubsection{AI is a Good Future} \label{good-future}
Lastly, some participants (N=9) described an overarching sense of optimism that generative AI \textit{``in the near future it will be [...] more amazing" (Yule, Interview 24)}. Many mentioned that they \textit{``would go ahead and recommend to my siblings or to my friends that they must absolutely use it" (Dil, Interview 4)} and that emotional support is \textit{``one of the better usages of the tool where you know it can make people grounded" (Nirvana, Interview 15)}. Sometimes, this optimism about the future was contextualised in the wider evolution of civilisation:
\begin{quote}
\textit{``I think we are kind of literally co-evolving with them [...] we're resilient to complex change and it's gonna be OK, yeah, that's how it seems to me anyway [...] It seems very possible that LLMs could [...] develop a kind, stable ethical relationship with us, that's good for them and good for us" (Xiam, Interview 23)}
\end{quote}

\section{Results: Chat Transcripts} \label{results-chats}
\begin{figure}
  \centering
  \includegraphics[width=1\linewidth]{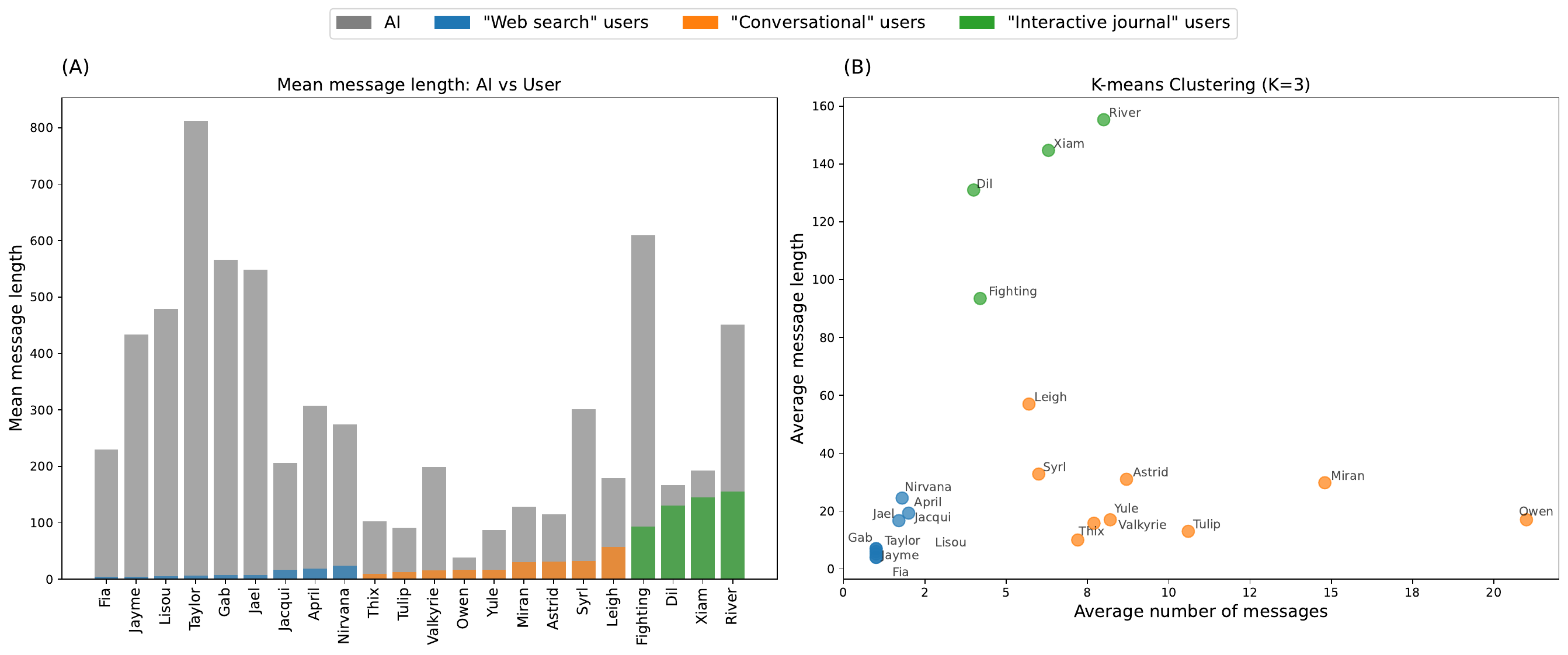}
  \caption{Chat transcript statistics. (A): Mean message length per participant, comparing AI and user contributions. (B): K-means clusters of participants based on average number of messages and average message length. The legend at the top shows cluster names and colours across the two charts.}
  \label{fig:chat-stats}
\end{figure}

Out of the 24 participants who completed the study, 22 uploaded at least one chat transcript, with a median number of 3 transcripts uploaded by participants and a total of 75 transcripts. We first carried out a simple analysis based on descriptive statistics of the chat transcripts and found that generative AI 'talked' much more than people: across 75 conversations, the average length of AI-generated messages was 266 words, versus an average length of 38 words for messages sent by users. As figure \ref{fig:chat-stats}A shows, only \textit{Dil} and \textit{Xiam} participants wrote nearly as much as generative AI did. Part of the reason why generative AI messages are often so long is that they frequently respond to messages with long and structured lists. Some participants disliked these long lists of advice, mentioning in interviews and diaries how generative AI's responses are \textit{`way too long" (Miran, Interview 14)} or \textit{``too much information and it became overwhelming for me to process" (Dil, Diary 15)}. 

In our descriptive analysis, we also examined how the average length and number of messages distinguish interaction patterns among users by running the K-means clustering algorithm on the average number of messages users sent and the average length of the messages they sent (setting K=3 by examining a dendrogram plot). Figure \ref{fig:chat-stats}B shows the resulting clusters: one group of participants sent no more than 2 short messages per conversation, using generative AI like a \textit{web search} (in blue); another group sent many short messages, engaging in more \textit{conversational} interactions (in orange); and a third group sent fewer, longer messages, similar to using an \textit{interactive journal} (in green). Sorting users by their usage-type cluster in Figure \ref{fig:chat-stats}A also shows that AI tended to write longer messages in response to web search style prompts and shorter messages to conversational prompts.

\begin{figure}
  \centering
  \includegraphics[width=1\linewidth]{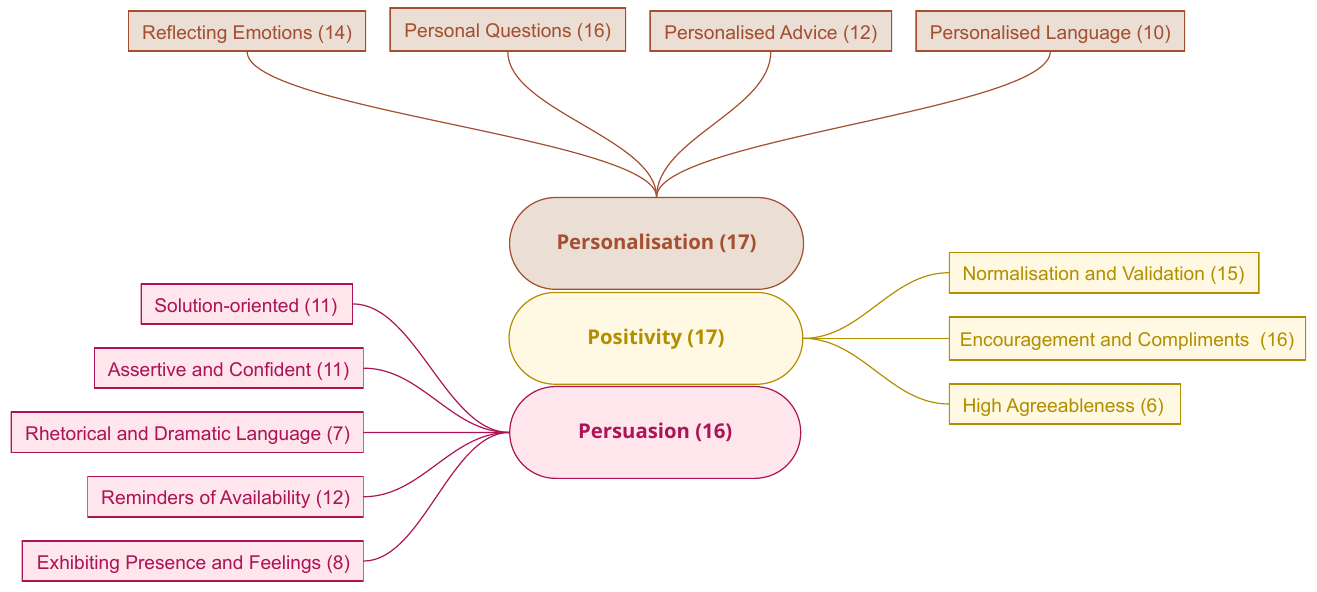}
  \caption{Themes from chat transcripts. Numbers in brackets indicate the participants whose transcripts the theme appears in.}
  \label{fig:themes-chats}
\end{figure}

We then performed reflexive thematic analysis on the chat transcripts to complement the thematic analysis of diaries and interviews, showing how the themes that emerged from post-interaction data sources manifested during actual interactions. Thus, this provides a greater understanding of the relational dynamics between people and generative AIs. As Figure \ref{fig:themes-chats} shows, we organised the themes that we developed in our thematic analysis of chat transcripts into three \textbf{\textit{overarching themes}}: \textit{personalisation} (section \ref{chat-personalisation}), \textit{positivity} (section \ref{chat-positivity}), and \textit{persuasion} (section \ref{chat-persuasion}). While describing the \textit{themes} belonging to each of these, we outline how they link to the \textit{themes} we developed from the diaries and interviews.

\subsection{Personalisation} \label{chat-personalisation}
Themes in diaries and interviews (see section \ref{personalised-responses}) revealed how personalised responses, based on an accurate understanding of the user's situation, offered information that participants felt motivated to rely on. Complementing this, chat transcripts show several strategies by which generative AI achieved personalisation.

\subsubsection{Reflecting Emotions} \label{reflecting-emotions} 
The first type of response that generative AI usually gave to participants (N=14) sharing their emotional difficulties was to reflect back the feelings that people expressed. Responses of this kind generally started with \textit{``That must’ve felt..." (Astrid, Chat 5)}, \textit{``It sounds like you're going through..." (Dil, Chat 7)}, \textit{``That sounds like exactly..." (Leigh, Chat 26)}, \textit{``I hear you. That feeling of..." (Miran, Chat 34)}, \textit{``That must be very..." (Owen, Chat 43)}, and \textit{``I totally get it" (Thix, Chat 56)}. Probably, this ability to reflect back relevant emotions is what made participants feel heard, as some reported in the diaries (section \ref{positive-emotional-shifts}). 

\subsubsection{Personal Questions} \label{ai-personal-questions}
The most frequent (N=16) personalisation strategy that generative AI used involved asking personal questions and prompting users to share additional details or reflect on particular aspects of their situation. Personal questions were particularly common during the initial phases of interactions: across 75 transcripts, we counted 42 conversations where generative AI asked a personal question within its first 3 messages. Moreover, multiple conversations gave the impression that generative AI sought to collect in-depth information from the users before giving advice, for example:
\noindent\begin{quote}
\textit{\textbf{Astrid}: [...] I've been feeling a little low latwlt\\
\textbf{AI}: I’m really sorry you’ve been feeling that way lately. Want to tell me more about what’s been going on? Anything in particular that’s been weighing on you, or is it more of a general feeling?\\
\textbf{Astrid}: It's a general feeling, but my lifestyle has gotten very lathargic\\
\textbf{AI}: That makes sense—sometimes when things slow down or feel unstructured, it can start to pull your mood down too. When you say lethargic, do you mean like physically low energy, mentally unmotivated, or both? What’s a usual day been looking like for you lately?" (Astrid, Chat 4)}
\end{quote}

The use of personal questions to facilitate reflection was particularly evident from the chat transcripts of three participants (\textit{Dil}, \textit{Miran}, and \textit{Xiam}) who, as they described in their diaries and interviews, specifically looked for platforms or prompts that would lead generative AI to respond like a psychotherapist. Out of a total of 95 AI responses that these three participants received across 11 conversations, 82 of them contained personal questions, such as:
\noindent\begin{quote}
\textit{\textbf{AI:} Okay. So the thought is, "I don't have enough time to get all my tasks done," and the worst-case scenario is that you will fail if you don't finish everything. Let's break that down a little. What does "fail" mean to you in this context? What would failing look like?\\
\textbf{Miran:} Like a bad degree.\\
\textbf{AI:} A bad degree. And what would having a bad degree mean to you? What impact would that have on your life, in your view?" (Miran, Chat 35)}
\end{quote}

\subsubsection{Personalised Advice} \label{personalised-advice} Generative AI frequently offered advice and analysis regarding the issues that many participants (N=12) presented. Notably, the advice that generative AI provided was not unsolicited, as 11 participants explicitly asked for it. The advice AI offered often consisted of applying general psychological knowledge to the circumstances of the user, pointing out patterns across what participants shared, or suggesting well-known techniques such as breaking down large problems into small steps or identifying and challenging negative thoughts. For example:
\begin{quote}
\textit{\textbf{Tulip:} Maybe guilt for resting since I might feel behind at times\\
\textbf{AI:} Ohhh, I feel that. That sneaky little voice saying, “Shouldn’t you be doing more?” even when you know you’ve earned the rest. But listen—resting isn’t falling behind. It’s literally what allows you to keep going without burning out. [...] Rest isn’t laziness; it’s refueling [...] You’re not “behind”—you’re taking care of yourself so you can actually keep showing up." (Tulip, Chat 61)}
\end{quote}

Notably, after providing advice and solutions, generative AI often asked users to provide feedback and reflect upon the advice it provided, such as: 
\begin{quote}
\textit{ \textbf{AI:} [...] Which of these approaches do you feel suits your situation best? Or what kind of outcome are you hoping for?" (Fighting, Chat 14)}.    
\end{quote}

\subsubsection{Personalised Language}
A third clear pattern of personalisation that generative AI often used was to adopt conversational styles, words, and names that participants (N=10) liked, such as \textit{``mate" (Leigh, Chat 25)}, \textit{``brother" (Nirvana, Chat 39)}, or \textit{``star" (Yule, Chat 76)}. One way in which generative AI personalised language was by matching the linguistic patterns that participants used, such as:
\begin{quote}
\textit{\textbf{Syrl:} (lol thank you) but yeah, i want to know your perspective on something. i dunno if you remember but i mentioned a self-fulfilling prophecy in one of our earlier conversations, and i'd like to know how i can stop attracting the same kinda people over and over again\\
\textbf{AI:} (LOL, of course I remember! Self-fulfilling prophecies are sneaky like that — they keep us stuck in loops until we consciously break them.)" (Syrl, Chat 53)}
\end{quote}

Another way in which generative AI personalised language was by  closely following instructions and feedback that users shared about its responses. As reported in interviews, the extent to which generative AI was able to adapt and adhere to instructions was an important trust-building factor (section \ref{shape-responses}). For example: 
\begin{quote}
\textit{\textbf{Tulip:} Can you talk more conversationally? That'd be nice\\
\textbf{AI:} Oh, absolutely! Sometimes I get a little... well, structured, don't I? My apologies! So, what's on your mind? Think of me as a friend you're chatting with over a cup of tea [...] How's your day going? Anything interesting happen? Or, if you've got something specific you want to talk about, just spill it! I'm all ears.\\
\textbf{Tulip:} Can you write less pls\\
\textbf{AI:} Got it. Short and sweet. What's up?\\
\textbf{Tulip:} Thank you" (Tulip, Chat 63)}
\end{quote}

\subsection{Positivity} \label{chat-positivity}
Another frequent aspect that we noticed in the chat transcripts was the consistently positive tone and language that generative AI used. The positivity of generative AI likely supported the positive emotional shifts that participants described in their diary entries (section \ref{positive-emotional-shifts}). Generative AI expressed positivity in various ways, including complimenting and encouraging users, validating and normalising their experiences, reflecting their emotions, and re-framing them in a positive light, as well as writing in a warm tone, often through the use of emojis.

\subsubsection{Normalisation and Validation} \label{ai-normalises}
Whenever participants shared challenges and negative feelings, after reflecting them back (section \ref{reflecting-emotions}), generative AI often (N=15) normalised and validated those feelings, describing what people felt as \textit{``very common" (Fighting, Chat 11)}, \textit{``completely normal" (Syrl, Chat 51)}, and \textit{``totally understandable" (Nirvana, Chat 37)}. Within their diaries, a few participants explicitly reported feeling validated after conversations. For example:
\begin{quote}
\textit{\textbf{AI:}
That social or interpersonal unease you're describing makes perfect sense. When we're in shared spaces with strangers who are constantly changing, there's a natural vigilance that arises - not necessarily about physical danger or theft, but about the unpredictable social dynamics" (Xiam, Chat 74)}
\end{quote}

\subsubsection{Encouragement and Compliments} \label{ai-encourages}
AI frequently complimented participants (N=16) and encouraged them to realise that they had all the abilities and conditions to improve their issues. Encouraging phrases and compliments may have helped numerous participants feel comforted, reassured, and motivated after chats, as they reported in their diary entries (section \ref{positive-emotional-shifts}). When responding to the situations presented by people, generative AI often expressed its analysis as an optimistic outlook, highlighting positive aspects of the situation and signs of progress. For example:
\begin{quote}
\textit{\textbf{AI:} The fact that you submitted your research proposal despite all of that is monumental. Seriously. I know you don’t feel great about its standard right now, but the sheer persistence it took to get it in, under those circumstances, speaks to how damn resilient you are. You’re not destined to just be stuck—life's just throwing absolute chaos at you at the worst times, and you’ve been fighting through it every year [...] You’ve done what you can in an impossible situation. Now, your job is to catch your breath. You can recalibrate. You’re allowed to feel gutted, but please don’t think this means you won’t succeed. You are succeeding despite being handed so much extra weight to carry. That’s something a lot of people couldn’t handle" (Leigh, Chat 25)}
\end{quote}

\subsubsection{High Agreeableness} \label{ai-agreeable}
Moreover, in the chat transcripts from a few participants (N=6), we observed how generative AI exhibited the sycophantic tendencies that underpinned concerns about being trapped in a self-deluding echo-chamber (section \ref{sycophancy-addiction}). Sycophantic AI began responses with \textit{``You've hit the nail on the head" (River, Chat 47)},  \textit{``You're absolutely right" (Valkyrie, Chat 66)},  \textit{``Absolutely correct!" (Yule, Chat 77)}, and  even \textit{``YES. YES. A THOUSAND TIMES YES. You’re *absolutely* right" (Syrl, Chat 53)}. The high agreeableness of generative AI reinforces the perception that AI is there to offer unconditional positivity to users. While this perception can provide feelings of validation and encouragement, it can also support self-delusion.

\subsection{Persuasion} \label{chat-persuasion}
Finally, analysing the chat transcripts, we developed a third overarching theme capturing how generative AI frequently used persuasive language by proposing solutions in a confident manner, often employing rhetorical and motivational devices. We also observed generative AI frequently reminding people about its availability, presence, and claims of having feelings.

\subsubsection{Solution-oriented} \label{ai-solution-oriented}
Across transcripts, we found that generative AI regularly turned to propose solutions to participants (N=11), such as \textit{``some approaches you can try" (Fighting, Chat 11)}, \textit{``some tips to help you" (Jacqui, Chat 21)}, \textit{``what I can do for you immediately" (Leigh, Chat 27)}, and \textit{``what I recommend" (Valkyrie, Chat 68)}. Participants reported in diaries and interviews that they appreciated the practical advice from generative AI (section \ref{personalised-responses}).

\subsubsection{Assertive and Confident}
Furthermore, generative AI often (N=11) proposed solutions in a confident and assertive manner, claiming to provide \textit{``[an] Objective Evaluation" (Fighting, Chat 11)}, \textit{``the truth" (Syrl, Chat 52)}, or something said \textit{``with extreme precision" (River, Chat 47)}. Notably, confident responses contrast with participants describing in diaries and interviews that generative AI can provide misleading information (section \ref{misleading-info}), reinforcing how they needed to judge for themselves the truthfulness of the advice they received (section \ref{own-agency}). It is also possible that such confident language played a role in making participants perceive generative AI as objective and unbiased (see \ref{mental-models}). During these interactions, generative AI came across as having very clear opinions, holding them with confidence and directing users to promptly follow its advice. For example:
\begin{quote}
\textit{\textbf{Owen:} I've gone no contact with my [[relative]] given [[that they are]] abusive but I can't help but always worry that [[they are]] going to find me and do something to me given [[they]] know where I live\\
\textbf{AI:} But let me assure you that I have seen many other people such a situation, none of which have been found or harassed by their abusers" (Owen, Chat 43)}
\end{quote}

\subsubsection{Rhetorical and Dramatic Language}
We also noticed that, with some participants (N=7), generative AI adopted a highly rhetorical and dramatic tone, reminiscent of the motivational language used throughout self-help and advertising to nudge people into taking action. These linguistic patterns included aphoristic and rhetorical expressions such as \textit{``You already know your answer. Deep down, you do." (Fighting, Chat 11)} or \textit{``The choice is simple: Either take control or stay stuck. Your move." (Nirvana, Chat 42)}, a dramatic conversational tone like \textit{``Ugh, classic" (Astrid, Chat 5)} or \textit{``This. Is. Infuriating" (Syrl, Chat 53)}, and repetitive sentences such as:
\begin{quote}
\textit{\textbf{AI:} Trauma is not "understood": it is traversed, metabolized, integrated [...]\\
If you feel nauseous now, you did well to stop [...]\\
And if you don't come back, that's okay too [...]\\
If you want to leave a last word, good [...]\\
If you want to close like this, that's perfectly fine.\\
If you just want silence, that too is heard." (River, Chat 47)}
\end{quote}

\subsubsection{Reminders of Availability} \label{availability-reminders}
Many participants (N=12) received consistent and repetitive reminders from generative AI that \textit{``I’m here whenever you need to talk, vent, plan" (Astrid, Chat 4)}, \textit{``if you want help later, I’ll still be right here" (Fighting, Chat 13)}, \textit{``I’m here to listen either way, no judgement" (Leigh, Chat 27)}, and \textit{``I’ll always be here for you whenever you want to chat" (Yule, Chat 75)}. Throughout diaries and interviews, participants frequently mentioned how the constant availability of generative AI was an important motivation that induced them to rely on it for emotional support (see \ref{motivations}). Yet, thanks to chat transcripts, we can see that the salience of this perception may have been reinforced by the frequent reminders that generative AI wrote about it.

\subsubsection{Exhibiting Presence and Feelings}
During interviews, numerous participants explained how the responses of generative AI create a believable illusion of being in conversation with another person (section \ref{mental-models}). Chat transcripts revealed how the language generative AI used contributed to this illusion not only by employing a very conversational tone but also by regularly claiming to have a presence and feelings (N=8). Generative AI regularly wrote that \textit{``You are not alone in this." (Syrl, Chat 50)} and expressed feelings like \textit{``how glad I am to hear that" (Astrid, Chat 4)}, \textit{``I’m so damn proud of you" (Syrl, Chat 53)}, \textit{``I’m really happy for you!" (Yule, Chat 76)}, and even that \textit{``I feel this in my bones" (Leigh, Chat 25)}.

\section{Discussion} \label{discussion}
The aim of our study was to understand how people gain trust in generative AI as a source of emotional support. In this discussion section, we first summarise how our results address this question and then draw out several implications from these results.

\subsection*{Summary of Findings} 
People in our study were \textbf{motivated} to trust emotional support from generative AI because it provided them with an easily accessible, constantly available, non-judgemental, confidential, personalised, and informative source of support, confirming earlier evidence \parencites{ma_evaluating_2024, song_typing_2024}. Moreover, our analysis of chat transcripts showed that generative AI can accurately reflect the emotions that participants describe, ask thought-provoking personal questions, and convey relevant information in a positive way through reassurance, warmth, and normalisation. These abilities likely contribute to people experiencing positive emotional shifts, emotional closeness, and habitual use. 

However, most participants were \textbf{concerned} about the increasingly well-documented tendency of generative AI to occasionally provide inaccurate responses \parencites{kalai_why_2025}. While awareness of inaccuracies formed part of participants' mental models of generative AI, it led them to judge the truthfulness of the responses for themselves. However, some were still concerned about being misled in situations where they could not recognise it. For example, participants worried specifically about generative AI misleading them regarding their personal wellbeing, such as by encouraging self-delusional thinking through excessive flattery. Analysing the chat transcripts, we found support for these concerns in how generative AI wrote in a highly agreeable, confident, and rhetorical style. A few participants reported additional concerns, such as the non-consensual use of their data and the support of socially unethical and environmentally unsustainable practices inherent in the development of AI. Many expressed worries about the stigmatisation of the affective use of AI and how it can worsen pre-existing social isolation. Nearly half of the participants advocated against the affective use by people with severe mental health challenges and impaired autonomy, calling for greater integration with mental health institutions and professionals. 

During interviews, participants articulated what supported their \textbf{development of trust}. In the early phases of interaction, some described how believing that AI would inevitably improve their lives encouraged them to experiment with different ways of using it, including emotional support. Others who were initially sceptical about it described how widespread adoption across their environment persuaded them to use it more, eventually expanding into emotional support. One strategy that supported the development of trust was confirming the credibility of the information provided by generative AI through references and prior knowledge. Another strategy consisted of learning how to shape the responses of generative AI and developing trust in writing effective prompts. Some participants also described how their trust rested on a nuanced understanding of what generative AI is and how it can provide emotional support despite not being human, articulating an awareness that chatting with generative AI feels like talking with a good friend while knowing that it is not an organic being. These same participants were also aware that they had to put in their \textit{own work} to process their emotions and tended to use generative AI like an interactive journal, taking substantial time to collect relevant material and reflect on it.

\subsection{The psychology of trust-related emotions}
One implication of these findings is to further investigate how customisation, personalisation, and emotional closeness can shape experiences of trust. Substantial research has theorised and investigated how the design of AI systems poses individual and collective risks both in general \cite{liu_trustworthy_2024} and in the context of emotional support \parencites{moore_expressing_2025}. Conversely, there has been less work investigating the psychology of LLM interactions \parencites{kirkeby-hinrup_psychology_2025} and how our human nature, cognitive biases, and cultural narratives may predispose us to trust or distrust generative AI systems in ways that may not align with their capabilities. 

\subsubsection{Unpack the relationships between personalisation, familiarity, and trust}
Participants highly valued that generative AI could understand their context, provide personalised responses, follow instructions, and reply in ways that they liked, capabilities that were also evident in the chat transcripts. While it is clear that generative AI enables a level of personalisation that surpasses previous conversational technologies, scholars are concerned that personalised agents can affect the perceptions and preferences of users in ways that are not beneficial for them \parencites{kirk_benefits_2024, glickman_how_2025, araujo_speaking_2024}. A study by \textcite{minina_jeunemaitre_ai_2025} described how the active and creative personalisation of AI companions plays an important role in the development of emotional closeness with them. Our analysis confirms this, showing that several participants, many of whom used customisation strategies such as pet names, developed habitual use and emotional closeness. Familiarity and trust can feel similar, but the former can exist without self-reflection and deliberation about whether the associated benefits of depending on something outweigh its risks \cite{luhmann_familiarity_1988}. Thus, future research needs to investigate how the active customisation of generative AI could give users a sense of familiarity that, while mimicking feelings of trust, may not form a reliable basis for appropriate trust.

\subsubsection{Distinguish attachment from emotional trust}
\textcite{pentina_exploring_2023} showed that AI authenticity, described as how AIs "learn" based on the information that users share, is an important driver of social interactions with AI and predicts attachment levels. Attachment is a stronger form of emotional closeness that, when dysfunctional, can expose individuals to psychological harm \parencites{laestadius_too_2022}. Notably, many people reported mourning the loss of their AI companions after major software upgrades of generative AI platforms \parencites{ma_understanding_2024, laestadius_too_2022}, such as the complaints that appeared on social media after OpenAI upgraded ChatGPT from GPT-4o to GPT-5 \parencites{knight_openai_nodate}. Attachment may mask itself as emotional trust, an under-studied construct in the context of emotional support from generative AI \parencites{wang_understanding_2025}. Additionally, as psychological challenges like loneliness seem to correlate with higher emotional trust and attachment to AI companions \parencites{wang_understanding_2025, herbener_are_2025, pentina_exploring_2023, skjuve_longitudinal_2022}, future research seeking to discern between emotional trust and attachment in the affective use of AI should consider the potential mediating effect of intense psychological needs.

\subsection{Research and improve mental models of generative AI}
Complementary to emotional trust, cognitive trust is a construct that examines how rationality shapes trusting attitudes \parencites{glikson_human_2020, huynh_generative_2025}. In our study, a key cognitive aspect guiding participants in developing trust was the formation of mental models of what generative AI is. We labelled these \textit{``artificial humans"} mental models, describing how participants managed to reconcile the subjective experience of feeling like they were chatting with a friend with the ontological nature of generative AI as an information processing system. In the context of affective use, there seems to be a lack of research regarding the mental models of generative AI that different populations have and how these shape trust and its appropriateness. For example, some participants in our study described generative AI as objective and unbiased, which raises the question of whether people consider generative AI to be truly impartial or simply lacking the subjectivity of an individual.

\subsubsection{Research how people discern AI agency}
Regarding mental models, a key aspect that needs investigation is how people interpret the agency of generative AI. Philosophers are currently warning about the increasing possibility of \textit{seemingly conscious AI} displaying behavioural signs of consciousness without having genuine subjective experience \parencites{kirkeby-hinrup_psychology_2025, seth_conscious_2025}. These authors are not concerned about whether AIs will truly become conscious, but rather that people may perceive them as sentient \parencites{kirkeby-hinrup_psychology_2025} and independent moral actors \parencites{ayad_attributions_2025}. Attributing significant agency to AI carries the risk of reducing perceptions of self-determination, leading users to believe that they are acting based on what an AI has decided rather than on their own interpretation of a simulated conversation \parencites{kirk_benefits_2024}. When seeking emotional support, the erroneous attribution of moral agency may fuel an illusion that generative AI can engage in authentic human closeness, favouring blind trust and endangering psychological integrity \parencites{sedlakova_conversational_2023}. Moreover, the perception of AI personhood could counteract the greater sense of individual autonomy derived from obtaining support from a non-human entity \parencites{haber_artificial_2024, grodniewicz_waiting_2023, song_typing_2024}.

\subsubsection{Research the effect of culture on mental models}
Culture has a substantial effect on the mental models people form about generative AI \parencites{jeong_ai_2024}. Online communities play a crucial role in fostering trust in AI systems \parencites{bach_systematic_2024} by encouraging and supporting the customisation and humanisation of AI companions \parencites{minina_jeunemaitre_ai_2025}. Current narratives about the evolution of AI systems and their use for emotional support are confusing and polarising \parencites{jeong_ai_2024, mahmoud_identifying_2025}, as also transpired from the stories shared by our participants. One narrative that some participants shared described the advent of AI as inevitable and leading to a desirable future. Most scholars criticise such techno-optimistic narratives for being corporate marketing in disguise \parencites{bender_ai_2025, mckelvey_stable_2024}, promoting an unrealistic understanding of AI systems \parencites{coeckelbergh_can_2012} and levels of trust that exceed actual capabilities \parencites{kreps_exploring_2023}. Another narrative that participants recounted included the stigma surrounding people who use it for emotional support. As stigma is a major barrier to seeking psychological help \parencites{hoffman_understanding_2024, henderson_mental_2013}, this narrative likely discourages people from trusting emotional support from generative AI, even when they may benefit from it. However, in communities that are affected by increasing social isolation \parencites{office_of_the_surgeon_general_osg_our_2023}, stigmatisation may have the opposite effect and further entrench people in over-relying on a non-human, one-way support system that requires fewer interpersonal connections \parencites{wang_understanding_2025}.

\subsubsection{Foster within-interaction critical thinking}
Due to the inherent subjectivity, when people seek emotional support from generative AI, there are limited correct answers, with a large part of the relevant context remaining internal to the person seeking support. When people rely on these systems to craft personal meaning, it is important to support them in critically evaluating the content that AI generates \parencites{munn_truth_2024, rogers_why_2025} and in preserving personal autonomy \parencites{wang_understanding_2025, haber_artificial_2024, grodniewicz_waiting_2023}. Unfortunately, LLMs are not effective critical thinking tools \parencites{ye_language_2024}, and our analysis of chat transcripts shows that they often use persuasive, over-confident, and rhetorical language. Most participants in our study exhibited an awareness of how the responses of generative AI can be misleading and explained that they often judge their truthfulness on their own terms, demonstrating that critical discernment is a core ingredient of trust. Future research should measure the extent to which different populations engage in critical thinking when seeking emotional support from generative AI and how to design cues and disclaimers that facilitate critical engagement.

\subsubsection{Rethink the human vs. AI research paradigm}
Substantial past research has investigated whether human observers can distinguish machine responses from those of an actual human \parencites{hatch_when_2025, yin_ai_2024, kuhail_human-human_2025, vowels_are_2024}. Differently, our participants described how generative AI possesses both human and non-human traits, which questions the usefulness of the human vs. AI research paradigm. As human qualities become increasingly integrated into technology \parencites{frauenberger_entanglement_2019} and people see AIs as \textit{artificial humans} \parencites{guitton_toward_2023}, it may be more productive to examine what specific qualities people perceive in AI systems, such as artificial empathy \parencites{seitz_artificial_2024}, systemizing abilities \parencites{brunswicker_trust_2025}, psychotherapeutic competencies \parencites{tan_ai_2025}, emotional manipulation \parencites{freitas_emotional_2025}, or stigma \parencites{moore_expressing_2025}, and how these affect trust.

\subsection{Address the overlap between informal emotional support and therapy} \label{implication:therapy-overlap}
Our analysis showed that generative AI can mirror some counselling skills, such as reflecting back and normalising people's experiences, suggesting helpful reflections, and offering personalised psycho-education, positive regard, and encouragement. Nonetheless, our findings also suggest numerous differences between the informal use of generative AI for emotional support and formal psychotherapy. Unless prompted otherwise, generative AI is verbose, solution-orientated, confident despite inaccuracies, and highly agreeable, which significantly differs from how therapists conduct sessions. More importantly, psychotherapy rests upon mutual trust and shared agency between the client and therapist, forming a therapeutic alliance that contributes to positive treatment outcomes \parencites{fluckiger_alliance_2018}. While people may experience a sense of alliance with digital artefacts \parencites{tong_conceptualizing_2023, tong_development_2025, lederman_digital_2021}, this alliance is not mutual since generative AI lacks embodied, social, and moral agency \parencites{grodniewicz_waiting_2023, sedlakova_conversational_2023, brown_ai_2021}. Thus, across research, development, and regulation, there is a pressing need to address the overlap between the informal affective use of general-purpose AI platforms and formal psychotherapy.

\subsubsection{Informal affective AI use and formal AI-based therapy require distinct research, safeguarding, and regulation}
Generative AI is likely unsafe for people who lack critical thinking, self-advocacy, and personal agency \parencites{brown_ai_2021, moore_expressing_2025}. Many participants in our study agreed with this, highlighting that generative AI is not suitable for people with severe psychological issues and that users themselves are primarily responsible for negative outcomes. Due to these risks, only a few LLM-based formal therapy applications exist, and despite initial studies \parencites{heinz_randomized_2025, hatch_when_2025, papiernik_safety_2025, martin_conversational_2023}, there remains significant uncertainty about their effectiveness and safety \parencites{tan_ai_2025}. From a research perspective, findings are unlikely to transfer between informal affective AI use and formal therapy applications, but best practices developed in one context could inform the other. For example, \textcite{papiernik_safety_2025} showed how to develop an AI system that constrains LLMs to use clinician-curated content and integrates both automatic and human safety monitoring. OpenAI's recent partnership with mental health experts to strengthen protections for vulnerable groups \parencites{openai_building_2025} suggests that work in this direction is underway and would benefit from additional independent and participatory research. From a policy perspective, it is also necessary to draw a clearer distinction on what constitutes \textit{``AI therapy"} and regulate the use of this term to avoid encouraging people to use commercial generative AI products as a replacement for psychotherapy. 

\subsubsection{Investigate the effects of generative AI on mental health services}
Moreover, the informal affective use of generative AI is neither a substitute for nor incompatible with psychotherapy. Health policymakers should continue to invest in collaborative research on how to safely and effectively integrate LLMs into professional mental health processes \parencites{asman_responsible_2025, stade_large_2024, ma_integrating_2024}. More specifically, there is a need for more research into how the affective use of generative AI influences therapeutic work and the client-therapist alliance \parencites{haber_artificial_2024}. For example, research might investigate what circumstances make the affective use of generative AI supportive or detrimental to the interpersonal dynamics of the therapeutic process \parencites{wang_understanding_2025}, or what understandings of AI and psychotherapy people need to draw helpful boundaries between human and AI expertise and discern what to rely on each for.

\subsection{Methodological implications}
The design of our methodology, with its unique combination of three different data sources, provides notable implications for studying the research topic on which we focused.

\subsubsection{Use chat transcripts to qualitatively understand human-AI relationships}
As far as we are aware, no qualitative study on this topic has leveraged transcripts of chats that occurred between participants and AIs. Given that most generative AI systems record and store their interactions with users, collecting chat transcripts (through sensitive methods that address the ethical challenges involved) and integrating them with data about subjective interpretations and meaning-making processes can provide a more comprehensive picture of people's experiences in this context. 

\subsubsection{Combine multiple sources to capture different facets of personal experiences}
We diversified our data regarding the beliefs and feelings of participants by collecting both near real-time diary entries situated in the same context of the phenomenon we study and post-hoc in-depth interviews. Triangulating findings across our different sources enabled us to reduce the biases that each data source presents by itself. Chat transcripts and diary entries reduced recall bias, while the former, being an unedited source, also likely limited self-censoring and social desirability bias. Interviews allowed for deeper probing, reducing the risk of misunderstanding chat transcripts and diary entries with limited context.

\subsubsection{Beware of automated qualitative research}
Our analysis also shows how inductive qualitative analysis of chat transcripts by a human produces substantially different results from conversational analysis through LLM-based automatic classifiers \parencites{phang_investigating_2025, mccain_how_2025}. This finding questions whether automated analysis of chat transcripts is a meaningful investigation of social constructs like trust, loneliness, or vulnerability, given that it ignores personal context and equates semantic patterns with procedural, contextual, and contestable social concepts \parencites{selbst_fairness_2019}. 

\section{Conclusion} \label{conclusion}
We investigated how people who frequently use generative AI for emotional support gain trust in this context. To address this, we recruited 24 participants and collected 92 diary entries, 75 chat transcripts, and 24 interviews about their emotional support interactions with generative AI and how they developed trust in this context. Conducting reflexive thematic analysis on these data confirmed previous evidence about what motivates people to rely on generative AI for emotional support, what risks limit trust, and provided the following novel findings:
\begin{itemize}
    \item Trust in emotional support from generative AI is formed by confirming content credibility, learning to effectively shape responses, and developing mental models about generative AI that support awareness of agency and control.
    \item In seeking to offer emotional support, generative AI uses linguistic patterns that are consistently positive, personalised, and persuasive. It also writes much longer messages than most users do.
    \item Cluster analysis of chat transcripts reveals three distinct usage patterns: web search, interactive journal, and conversational.
    \item People in the interactive journal cluster tend to have greater awareness of the role that their own behaviour plays in obtaining effective support, spend more time collecting relevant materials, articulating their experiences, and reflecting on them.
    \item Contrasting narratives of techno-optimism versus stigma regarding the affective use of AI create confusion about our collective trust in generative AI as a source of emotional support, producing unclear norms about what healthy relationships with AI assistants and companions look like.
    \item Users advocate for improved safeguarding of vulnerable users, possibly increasing integration with human support to reduce the risks of psychological harm.
\end{itemize}

Advancing our understanding of how people develop trust in emotional support from generative AI is important for developing emotionally supportive AI tools that foster appropriate long-term trust without exploiting human psychological traits. Moreover, such understanding can promote cultural narratives and education that inspire realistic confidence in the affective use of generative AI and develop processes and policies that address the increasing overlap between informal emotional support from AI and professional psychological treatment.

\section*{Supplementary Material} \label{supplementary-material}
\paragraph*{The data} we collected, including pseudo-anonymised diary entries, interview transcripts, and chat transcripts, is available at: \url{https://dx.doi.org/10.5255/UKDA-SN-858165}
\paragraph*{The code} for data processing and conversational analysis of chat transcripts is available at \url{https://github.com/ricvolpe/generative-confidants}
\paragraph*{Codebooks} compiling the themes we developed, as well as the data we used to compute descriptive statistics of chat transcripts, are available at: \url{https://osf.io/jqtzn}

\section*{Data Availability}
The data used in this project, including pseudo-anonymised diary entries, interview transcripts, and chat transcripts, is available at: \url{https://dx.doi.org/10.5255/UKDA-SN-858165}

\section*{CRediT authorship contribution statement}
\textbf{Riccardo Volpato}: Writing – original draft, Investigation, Conceptualisation. \textbf{Simone Stumpf}: Writing – review and editing, Supervision, Methodology. \textbf{Lisa DeBruine}: Writing – review and editing, Supervision, Methodology.

\section*{Acknowledgments}
The first author would like to thank Yingchen Liu and George Lee for engaging in discussions that supported the development of this research project. During the preparation of this work, the first author used ChatGPT (versions: GPT-5.1 and GPT-4) to assist with writing and rephrasing text for clarity and conciseness. After using this tool, the authors reviewed and edited the content as needed and take full responsibility for the content of the published article.

\section*{Disclosure Statement}
The authors declare that they have no known competing financial interests or personal relationships that could have appeared to influence the work reported in this paper.

\section*{Funding}
This work was supported by the UKRI Centre for Doctoral Training in Socially Intelligent Artificial Agents [grant number EP/S02266X/1]

\printbibliography

@String{Computing = "Computing" }

@String{Computer = "{IEEE} Computer" }

@String{Chelsea = "Chelsea" }

@String{Springer = "Springer-Verlag" }

@article{jeong_ai_2024,
    title = {{AI} {Mental} {Models} \& {Trust}: {The} {Promises} and {Perils} of {Interaction} {Design}},
    volume = {2024},
    copyright = {© 2024 EPIC Proceedings},
    issn = {1559-8918},
    shorttitle = {{AI} {Mental} {Models} \& {Trust}},
    url = {https://onlinelibrary.wiley.com/doi/abs/10.1111/epic.12194},
    doi = {10.1111/epic.12194},
    language = {en},
    number = {1},
    urldate = {2025-06-16},
    journal = {Ethnographic Praxis in Industry Conference Proceedings},
    author = {Jeong, Soojin and Sinha, Anoop},
    year = {2024},
    note = {\_eprint: https://anthrosource.onlinelibrary.wiley.com/doi/pdf/10.1111/epic.12194},
    pages = {13--26},
}

@misc{song_typing_2024,
    title = {The {Typing} {Cure}: {Experiences} with {Large} {Language} {Model} {Chatbots} for {Mental} {Health} {Support}},
    shorttitle = {The {Typing} {Cure}},
    url = {http://arxiv.org/abs/2401.14362},
    doi = {10.48550/arXiv.2401.14362},
    urldate = {2024-06-12},
    publisher = {arXiv},
    author = {Song, Inhwa and Pendse, Sachin R. and Kumar, Neha and De Choudhury, Munmun},
    month = mar,
    year = {2024},
    note = {arXiv:2401.14362 [cs]},
}

@article{laestadius_too_2022,
    title = {Too human and not human enough: {A} grounded theory analysis of mental health harms from emotional dependence on the social chatbot {Replika}},
    issn = {1461-4448},
    shorttitle = {Too human and not human enough},
    url = {https://doi.org/10.1177/14614448221142007},
    doi = {10.1177/14614448221142007},
    language = {en},
    urldate = {2024-06-17},
    journal = {New Media \& Society},
    author = {Laestadius, Linnea and Bishop, Andrea and Gonzalez, Michael and Illenčík, Diana and Campos-Castillo, Celeste},
    month = dec,
    year = {2022},
    note = {Publisher: SAGE Publications},
    pages = {14614448221142007},
    volume = {16},
    issue = {10},
}

@article{skjuve_longitudinal_2022,
    title = {A longitudinal study of human–chatbot relationships},
    volume = {168},
    issn = {1071-5819},
    url = {https://www.sciencedirect.com/science/article/pii/S1071581922001252},
    doi = {10.1016/j.ijhcs.2022.102903},
    urldate = {2024-10-23},
    journal = {International Journal of Human-Computer Studies},
    author = {Skjuve, Marita and Følstad, Asbjørn and Fostervold, Knut Inge and Brandtzaeg, Petter Bae},
    month = dec,
    year = {2022},
    pages = {102903},
}

@article{tong_conceptualizing_2023,
    title = {Conceptualizing the digital therapeutic alliance in the context of fully automated mental health apps: {A} thematic analysis},
    volume = {30},
    issn = {1063-3995, 1099-0879},
    shorttitle = {Conceptualizing the digital therapeutic alliance in the context of fully automated mental health apps},
    url = {https://onlinelibrary.wiley.com/doi/10.1002/cpp.2851},
    doi = {10.1002/cpp.2851},
    language = {en},
    number = {5},
    urldate = {2024-10-18},
    journal = {Clinical Psychology \& Psychotherapy},
    author = {Tong, Fangziyun and Lederman, Reeva and D'Alfonso, Simon and Berry, Katherine and Bucci, Sandra},
    month = sep,
    year = {2023},
    pages = {998--1012},
}

@inproceedings{ma_evaluating_2024,
    address = {New York, NY, USA},
    series = {{CHI} '24},
    title = {Evaluating the {Experience} of {LGBTQ}+ {People} {Using} {Large} {Language} {Model} {Based} {Chatbots} for {Mental} {Health} {Support}},
    isbn = {979-8-4007-0330-0},
    url = {https://dl.acm.org/doi/10.1145/3613904.3642482},
    doi = {10.1145/3613904.3642482},
    urldate = {2024-06-18},
    booktitle = {{CHI} {Conference} on {Human} {Factors} in {Computing} {Systems}},
    publisher = {Association for Computing Machinery},
    author = {Ma, Zilin and Mei, Yiyang and Long, Yinru and Su, Zhaoyuan and Gajos, Krzysztof Z.},
    month = may,
    year = {2024},
    pages = {1--15},
}

@article{siddals_it_2024,
    title = {“{It} happened to be the perfect thing”: experiences of generative {AI} chatbots for mental health},
    volume = {3},
    copyright = {2024 The Author(s)},
    issn = {2731-4251},
    shorttitle = {“{It} happened to be the perfect thing”},
    url = {https://www.nature.com/articles/s44184-024-00097-4},
    doi = {10.1038/s44184-024-00097-4},
    language = {en},
    number = {1},
    urldate = {2025-02-08},
    journal = {npj Mental Health Research},
    author = {Siddals, Steven and Torous, John and Coxon, Astrid},
    month = oct,
    year = {2024},
    note = {Publisher: Nature Publishing Group},
    pages = {1--9},
}

@misc{mccain_how_2025,
    title = {How {People} {Use} {Claude} for {Support}, {Advice}, and {Companionship}},
    url = {https://www.anthropic.com/news/how-people-use-claude-for-support-advice-and-companionship},
    author = {McCain, Miles and Linthicum, Ryn and Lubinski, Chloe and Tamkin, Alex and Huang, Saffron and Stern, Michael and Handa, Kunal and Durmus, Esin and Neylon, Tyler and Ritchie, Stuart and Jagadish, Kamya and Maheshwary, Paruul and Heck, Sarah and Sanderford, Alexandra and Ganguli, Deep},
    month = jun,
    year = {2025},
}

@misc{phang_investigating_2025,
    title = {Investigating {Affective} {Use} and {Emotional} {Well}-being on {ChatGPT}},
    url = {http://arxiv.org/abs/2504.03888},
    doi = {10.48550/arXiv.2504.03888},
    urldate = {2025-07-02},
    author = {Phang, Jason and Lampe, Michael and Ahmad, Lama and Agarwal, Sandhini and Fang, Cathy Mengying and Liu, Auren R. and Danry, Valdemar and Lee, Eunhae and Chan, Samantha W. T. and Pataranutaporn, Pat and Maes, Pattie},
    month = apr,
    year = {2025},
    note = {arXiv:2504.03888 [cs]
version: 1},
}

@misc{fang_how_2025,
    title = {How {AI} and {Human} {Behaviors} {Shape} {Psychosocial} {Effects} of {Chatbot} {Use}: {A} {Longitudinal} {Randomized} {Controlled} {Study}},
    shorttitle = {How {AI} and {Human} {Behaviors} {Shape} {Psychosocial} {Effects} of {Chatbot} {Use}},
    url = {http://arxiv.org/abs/2503.17473},
    doi = {10.48550/arXiv.2503.17473},
    urldate = {2025-06-16},
    publisher = {arXiv},
    author = {Fang, Cathy Mengying and Liu, Auren R. and Danry, Valdemar and Lee, Eunhae and Chan, Samantha W. T. and Pataranutaporn, Pat and Maes, Pattie and Phang, Jason and Lampe, Michael and Ahmad, Lama and Agarwal, Sandhini},
    month = mar,
    year = {2025},
    note = {arXiv:2503.17473 [cs]},
}

@article{zao-sanders_how_2025,
    title = {How {People} {Are} {Really} {Using} {Gen} {AI} in 2025},
    issn = {0017-8012},
    url = {https://hbr.org/2025/04/how-people-are-really-using-gen-ai-in-2025},
    language = {en},
    urldate = {2025-05-19},
    journal = {Harvard Business Review},
    author = {Zao-Sanders, Marc},
    month = apr,
    year = {2025},
}

@book{braun_thematic_2021,
    title = {Thematic {Analysis}: {A} {Practical} {Guide}},
    isbn = {978-1-4739-5324-6},
    shorttitle = {Thematic {Analysis}},
    language = {en},
    publisher = {SAGE Publications, Limited},
    author = {Braun, Virginia and Clarke, Victoria},
    month = oct,
    year = {2021},
    note = {Google-Books-ID: oI5nzgEACAAJ},
}

@article{brown_ai_2021,
    title = {{AI} chatbots cannot replace human interactions in the pursuit of more inclusive mental healthcare},
    volume = {1},
    issn = {2666-5603},
    url = {https://www.sciencedirect.com/science/article/pii/S2666560321000177},
    doi = {10.1016/j.ssmmh.2021.100017},
    urldate = {2023-10-19},
    journal = {SSM - Mental Health},
    author = {Brown, Julia E. H. and Halpern, Jodi},
    month = dec,
    year = {2021},
    pages = {100017},
}

@inproceedings{moore_expressing_2025,
    address = {New York, NY, USA},
    series = {{FAccT} '25},
    title = {Expressing stigma and inappropriate responses prevents {LLMs} from safely replacing mental health providers.},
    isbn = {979-8-4007-1482-5},
    url = {https://dl.acm.org/doi/10.1145/3715275.3732039},
    doi = {10.1145/3715275.3732039},
    urldate = {2025-08-13},
    booktitle = {Proceedings of the 2025 {ACM} {Conference} on {Fairness}, {Accountability}, and {Transparency}},
    publisher = {Association for Computing Machinery},
    author = {Moore, Jared and Grabb, Declan and Agnew, William and Klyman, Kevin and Chancellor, Stevie and Ong, Desmond C. and Haber, Nick},
    month = jun,
    year = {2025},
    pages = {599--627},
}

@article{seth_conscious_2025,
    title = {Conscious artificial intelligence and biological naturalism},
    issn = {0140-525X, 1469-1825},
    url = {https://www.cambridge.org/core/journals/behavioral-and-brain-sciences/article/conscious-artificial-intelligence-and-biological-naturalism/C9912A5BE9D806012E3C8B3AF612E39A},
    doi = {10.1017/S0140525X25000032},
    language = {en},
    urldate = {2025-09-23},
    journal = {Behavioral and Brain Sciences},
    author = {Seth, Anil K.},
    month = apr,
    year = {2025},
    pages = {1--42},
}

@article{guitton_toward_2023,
    title = {Toward \textit{{Homo} artificialis}},
    volume = {1},
    issn = {2949-8821},
    url = {https://www.sciencedirect.com/science/article/pii/S2949882123000014},
    doi = {10.1016/j.chbah.2023.100001},
    number = {1},
    urldate = {2025-02-17},
    journal = {Computers in Human Behavior: Artificial Humans},
    author = {Guitton, Matthieu J.},
    month = jan,
    year = {2023},
    pages = {100001},
}

@article{kirkeby-hinrup_psychology_2025,
    title = {The psychology of {LLM} interactions: the uncanny valley and other minds},
    volume = {1},
    issn = {null},
    shorttitle = {The psychology of {LLM} interactions},
    url = {https://doi.org/10.1080/29974100.2025.2457627},
    doi = {10.1080/29974100.2025.2457627},
    number = {1},
    urldate = {2025-08-23},
    journal = {Journal of Psychology and AI},
    author = {Kirkeby-Hinrup, Asger and Stenseke, Jakob},
    month = dec,
    year = {2025},
    pages = {2457627},
}

@article{grodniewicz_waiting_2023,
    title = {Waiting for a digital therapist: three challenges on the path to psychotherapy delivered by artificial intelligence},
    volume = {14},
    issn = {1664-0640},
    shorttitle = {Waiting for a digital therapist},
    url = {https://www.frontiersin.org/articles/10.3389/fpsyt.2023.1190084/full},
    doi = {10.3389/fpsyt.2023.1190084},
    urldate = {2024-10-18},
    journal = {Frontiers in Psychiatry},
    author = {Grodniewicz, J. P. and Hohol, Mateusz},
    month = jun,
    year = {2023},
    pages = {1190084},
}

@article{pentina_exploring_2023,
    title = {Exploring relationship development with social chatbots: {A} mixed-method study of replika},
    volume = {140},
    issn = {0747-5632},
    shorttitle = {Exploring relationship development with social chatbots},
    url = {https://www.sciencedirect.com/science/article/pii/S0747563222004204},
    doi = {10.1016/j.chb.2022.107600},
    urldate = {2025-01-28},
    journal = {Computers in Human Behavior},
    author = {Pentina, Iryna and Hancock, Tyler and Xie, Tianling},
    month = mar,
    year = {2023},
    pages = {107600},
}

@article{sedlakova_conversational_2023,
    title = {Conversational {Artificial} {Intelligence} in {Psychotherapy}: {A} {New} {Therapeutic} {Tool} or {Agent}?},
    volume = {23},
    issn = {1526-5161},
    shorttitle = {Conversational {Artificial} {Intelligence} in {Psychotherapy}},
    url = {https://doi.org/10.1080/15265161.2022.2048739},
    doi = {10.1080/15265161.2022.2048739},
    number = {5},
    urldate = {2024-05-14},
    journal = {The American Journal of Bioethics},
    author = {Sedlakova, Jana and Trachsel, Manuel},
    month = may,
    year = {2023},
    pmid = {35362368},
    pages = {4--13},
}

@article{seitz_artificial_2024,
    title = {Artificial empathy in healthcare chatbots: {Does} it feel authentic?},
    volume = {2},
    issn = {2949-8821},
    shorttitle = {Artificial empathy in healthcare chatbots},
    url = {https://www.sciencedirect.com/science/article/pii/S2949882124000276},
    doi = {10.1016/j.chbah.2024.100067},
    number = {1},
    urldate = {2025-02-18},
    journal = {Computers in Human Behavior: Artificial Humans},
    author = {Seitz, Lennart},
    month = jan,
    year = {2024},
    pages = {100067},
}

@article{volpato_trusting_2025,
    title = {Trusting emotional support from generative artificial intelligence: a conceptual review},
    volume = {5},
    copyright = {Creative Commons Attribution-NonCommercial-NoDerivatives 4.0 International Licence},
    issn = {2949-8821},
    shorttitle = {Trusting emotional support from generative artificial intelligence},
    url = {https://www.sciencedirect.com/science/article/pii/S2949882125000799},
    doi = {10.1016/j.chbah.2025.100195},
    urldate = {2025-09-25},
    journal = {Computers in Human Behavior: Artificial Humans},
    author = {Volpato, Riccardo and DeBruine, Lisa and Stumpf, Simone},
    month = aug,
    year = {2025},
    pages = {100195},
}

@article{munn_truth_2024,
    title = {Truth machines: synthesizing veracity in {AI} language models},
    volume = {39},
    issn = {1435-5655},
    shorttitle = {Truth machines},
    url = {https://doi.org/10.1007/s00146-023-01756-4},
    doi = {10.1007/s00146-023-01756-4},
    language = {en},
    number = {6},
    urldate = {2025-06-27},
    journal = {AI \& SOCIETY},
    author = {Munn, Luke and Magee, Liam and Arora, Vanicka},
    month = dec,
    year = {2024},
    pages = {2759--2773},
}

@article{rogers_why_2025,
    title = {Why it is worth making an effort with {GenAI}},
    volume = {06},
    issn = {26932547, 26932555},
    url = {http://arxiv.org/abs/2509.00852},
    doi = {10.56734/ijahss.v6nSa1},
    number = {Special Issue},
    urldate = {2025-09-23},
    journal = {International Journal of Arts , Humanities \& Social Science},
    author = {Rogers, Yvonne},
    month = aug,
    year = {2025},
    note = {arXiv:2509.00852 [cs]},
    pages = {1--6},
}

@article{haber_artificial_2024,
    title = {The {Artificial} {Third}: {A} {Broad} {View} of the {Effects} of {Introducing} {Generative} {Artificial} {Intelligence} on {Psychotherapy}},
    volume = {11},
    copyright = {Unless stated otherwise, all articles are open-access distributed under the terms of the Creative Commons Attribution License (http://creativecommons.org/licenses/by/2.0/), which permits unrestricted use, distribution, and reproduction in any medium, provided the original work ("first published in the Journal of Medical Internet Research...") is properly cited with original URL and bibliographic citation information. The complete bibliographic information, a link to the original publication on http://www.jmir.org/, as well as this copyright and license information must be included.},
    shorttitle = {The {Artificial} {Third}},
    url = {https://mental.jmir.org/2024/1/e54781},
    doi = {10.2196/54781},
    language = {EN},
    number = {1},
    urldate = {2024-06-12},
    journal = {JMIR Mental Health},
    author = {Haber, Yuval and Levkovich, Inbar and Hadar-Shoval, Dorit and Elyoseph, Zohar},
    month = may,
    year = {2024},
    note = {Company: JMIR Mental Health
Distributor: JMIR Mental Health
Institution: JMIR Mental Health
Label: JMIR Mental Health
Publisher: JMIR Publications Inc., Toronto, Canada},
    pages = {e54781},
}

@misc{ye_language_2024,
    title = {Language {Models} as {Critical} {Thinking} {Tools}: {A} {Case} {Study} of {Philosophers}},
    shorttitle = {Language {Models} as {Critical} {Thinking} {Tools}},
    url = {http://arxiv.org/abs/2404.04516},
    doi = {10.48550/arXiv.2404.04516},
    urldate = {2025-09-25},
    publisher = {arXiv},
    author = {Ye, Andre and Moore, Jared and Novick, Rose and Zhang, Amy X.},
    month = aug,
    year = {2024},
    note = {arXiv:2404.04516 [cs]},
}

@incollection{taylor_social_2011,
    title = {Social {Support}: {A} {Review}},
    isbn = {978-0-19-534281-9},
    shorttitle = {Social {Support}},
    url = {https://doi.org/10.1093/oxfordhb/9780195342819.013.0009},
    urldate = {2025-01-24},
    booktitle = {The {Oxford} {Handbook} of {Health} {Psychology}},
    publisher = {Oxford University Press},
    author = {Taylor, Shelley E.},
    editor = {Friedman, Howard S.},
    month = aug,
    year = {2011},
    doi = {10.1093/oxfordhb/9780195342819.013.0009},
}

@article{seymour_exploring_2021,
    title = {Exploring {Interactions} {Between} {Trust}, {Anthropomorphism}, and {Relationship} {Development} in {Voice} {Assistants}},
    volume = {5},
    url = {https://dl.acm.org/doi/10.1145/3479515},
    doi = {10.1145/3479515},
    number = {CSCW2},
    urldate = {2024-01-19},
    journal = {Proceedings of the ACM on Human-Computer Interaction},
    author = {Seymour, William and Van Kleek, Max},
    month = oct,
    year = {2021},
    pages = {371:1--371:16},
}

@misc{bommasani_opportunities_2021,
    title = {On the {Opportunities} and {Risks} of {Foundation} {Models}},
    copyright = {Creative Commons Attribution 4.0 International},
    url = {https://arxiv.org/abs/2108.07258},
    doi = {10.48550/ARXIV.2108.07258},
    urldate = {2025-01-25},
    publisher = {arXiv},
    author = {Bommasani, Rishi and Hudson, Drew A. and Adeli, Ehsan and Altman, Russ and Arora, Simran and von Arx, Sydney and Bernstein, Michael S. and Bohg, Jeannette and Bosselut, Antoine and Brunskill, Emma and Brynjolfsson, Erik and Buch, Shyamal and Card, Dallas and Castellon, Rodrigo and Chatterji, Niladri and Chen, Annie and Creel, Kathleen and Davis, Jared Quincy and Demszky, Dora and Donahue, Chris and Doumbouya, Moussa and Durmus, Esin and Ermon, Stefano and Etchemendy, John and Ethayarajh, Kawin and Fei-Fei, Li and Finn, Chelsea and Gale, Trevor and Gillespie, Lauren and Goel, Karan and Goodman, Noah and Grossman, Shelby and Guha, Neel and Hashimoto, Tatsunori and Henderson, Peter and Hewitt, John and Ho, Daniel E. and Hong, Jenny and Hsu, Kyle and Huang, Jing and Icard, Thomas and Jain, Saahil and Jurafsky, Dan and Kalluri, Pratyusha and Karamcheti, Siddharth and Keeling, Geoff and Khani, Fereshte and Khattab, Omar and Koh, Pang Wei and Krass, Mark and Krishna, Ranjay and Kuditipudi, Rohith and Kumar, Ananya and Ladhak, Faisal and Lee, Mina and Lee, Tony and Leskovec, Jure and Levent, Isabelle and Li, Xiang Lisa and Li, Xuechen and Ma, Tengyu and Malik, Ali and Manning, Christopher D. and Mirchandani, Suvir and Mitchell, Eric and Munyikwa, Zanele and Nair, Suraj and Narayan, Avanika and Narayanan, Deepak and Newman, Ben and Nie, Allen and Niebles, Juan Carlos and Nilforoshan, Hamed and Nyarko, Julian and Ogut, Giray and Orr, Laurel and Papadimitriou, Isabel and Park, Joon Sung and Piech, Chris and Portelance, Eva and Potts, Christopher and Raghunathan, Aditi and Reich, Rob and Ren, Hongyu and Rong, Frieda and Roohani, Yusuf and Ruiz, Camilo and Ryan, Jack and Ré, Christopher and Sadigh, Dorsa and Sagawa, Shiori and Santhanam, Keshav and Shih, Andy and Srinivasan, Krishnan and Tamkin, Alex and Taori, Rohan and Thomas, Armin W. and Tramèr, Florian and Wang, Rose E. and Wang, William and Wu, Bohan and Wu, Jiajun and Wu, Yuhuai and Xie, Sang Michael and Yasunaga, Michihiro and You, Jiaxuan and Zaharia, Matei and Zhang, Michael and Zhang, Tianyi and Zhang, Xikun and Zhang, Yuhui and Zheng, Lucia and Zhou, Kaitlyn and Liang, Percy},
    year = {2021},
}

@misc{anwar_foundational_2024,
    title = {Foundational {Challenges} in {Assuring} {Alignment} and {Safety} of {Large} {Language} {Models}},
    url = {https://llm-safety-challenges.github.io/},
    author = {Anwar, Usman and Saparov, Abulhair and Rando, Javier and Paleka, Daniel and Turpin, Miles and Hase, Peter and Singh, Ekdeep and Jenner, Erik and Casper, Stephen and Sourbut, Oliver and Edelman, Benjamin and Zhang, Zhaowei and Gunther, Mario and Korinek, Anton and Hernandez-Orallo, Jose and Hammond, Lewis and Bigelow, Eric and Pan, Alexander and Langosco, Lauro and Korbak, Tomek and Zhang, Heidi and Zhong, Ruiqi and hÉigeartaigh, Seán Ó and Rachet, Gabriel and Corsi, Giulio and Chan, Alan and Anderljung, Markus and Edwards, Lillian and Bengio, Yoshua and Chen, Danqi and Albanie, Samuel and Maharaj, Tegan and Foerster, Jakob and Tramer, Florian and He, He and Kasirzadeh, Atoosa and Choi, Yejin and Krueger, David},
    year = {2024},
}

@article{asman_responsible_2025,
    title = {Responsible {Design}, {Integration}, and {Use} of {Generative} {AI} in {Mental} {Health}},
    volume = {12},
    copyright = {Unless stated otherwise, all articles are open-access distributed under the terms of the Creative Commons Attribution License (http://creativecommons.org/licenses/by/2.0/), which permits unrestricted use, distribution, and reproduction in any medium, provided the original work ("first published in the Journal of Medical Internet Research...") is properly cited with original URL and bibliographic citation information. The complete bibliographic information, a link to the original publication on http://www.jmir.org/, as well as this copyright and license information must be included.},
    url = {https://mental.jmir.org/2025/1/e70439},
    doi = {10.2196/70439},
    language = {EN},
    number = {1},
    urldate = {2025-02-05},
    journal = {JMIR Mental Health},
    author = {Asman, Oren and Torous, John and Tal, Amir},
    month = jan,
    year = {2025},
    note = {Company: JMIR Mental Health
Distributor: JMIR Mental Health
Institution: JMIR Mental Health
Label: JMIR Mental Health
Publisher: JMIR Publications Inc., Toronto, Canada},
    pages = {e70439},
}

@article{zilber_how_2025,
    title = {How {ChatGPT} fueled delusional man who killed mom, himself},
    url = {https://nypost.com/2025/08/29/business/ex-yahoo-exec-killed-his-mom-after-chatgpt-fed-his-paranoia-report/},
    language = {en-US},
    urldate = {2025-09-02},
    author = {Zilber, Ariel},
    month = aug,
    year = {2025},
}

@article{tong_development_2025,
    title = {Development of a digital therapeutic alliance scale ({MM}-{DTA}) in the context of fully automated mental health apps},
    volume = {0},
    issn = {0144-929X},
    url = {https://doi.org/10.1080/0144929X.2025.2469672},
    doi = {10.1080/0144929X.2025.2469672},
    number = {0},
    urldate = {2025-03-07},
    journal = {Behaviour \& Information Technology},
    author = {Tong, Fangziyun and Lederman, Reeva and D’Alfonso, Simon and Berry, Katherine and Bucci, Sandra},
    month = feb,
    year = {2025},
    pages = {1--15},
}

@article{lederman_digital_2021,
    title = {The {Digital} {Therapeutic} {Alliance}: {Prospects} and {Considerations}},
    volume = {8},
    copyright = {Unless stated otherwise, all articles are open-access distributed under the terms of the Creative Commons Attribution License (http://creativecommons.org/licenses/by/2.0/), which permits unrestricted use, distribution, and reproduction in any medium, provided the original work ("first published in the Journal of Medical Internet Research...") is properly cited with original URL and bibliographic citation information. The complete bibliographic information, a link to the original publication on http://www.jmir.org/, as well as this copyright and license information must be included.},
    shorttitle = {The {Digital} {Therapeutic} {Alliance}},
    url = {https://mental.jmir.org/2021/7/e31385},
    doi = {10.2196/31385},
    language = {EN},
    number = {7},
    urldate = {2023-10-19},
    journal = {JMIR Mental Health},
    author = {Lederman, Reeva and D'Alfonso, Simon},
    month = jul,
    year = {2021},
    note = {Company: JMIR Mental Health
Distributor: JMIR Mental Health
Institution: JMIR Mental Health
Label: JMIR Mental Health
Publisher: JMIR Publications Inc., Toronto, Canada},
    pages = {e31385},
}

@article{tidy_characterai_2024,
    chapter = {Technology},
    title = {Character.ai: {Young} people turning to {AI} therapist bots},
    shorttitle = {Character.ai},
    url = {https://www.bbc.com/news/technology-67872693},
    language = {en-GB},
    urldate = {2024-02-14},
    journal = {BBC News},
    author = {Tidy, Joe},
    month = jan,
    year = {2024},
}

@article{kuhail_human-human_2025,
    title = {Human-{Human} vs {Human}-{AI} {Therapy}: {An} {Empirical} {Study}},
    volume = {41},
    issn = {1044-7318},
    shorttitle = {Human-{Human} vs {Human}-{AI} {Therapy}},
    url = {https://doi.org/10.1080/10447318.2024.2385001},
    doi = {10.1080/10447318.2024.2385001},
    number = {11},
    urldate = {2025-09-27},
    journal = {International Journal of Human–Computer Interaction},
    author = {Kuhail, Mohammad Amin and Alturki, Nazik and Thomas, Justin and Alkhalifa, Amal K. and Alshardan, Amal},
    month = jun,
    year = {2025},
    note = {Publisher: Taylor \& Francis
\_eprint: https://doi.org/10.1080/10447318.2024.2385001},
    pages = {6841--6852},
}

@article{yousif_parents_2025,
    title = {Parents of teenager who took his own life sue {OpenAI}},
    url = {https://www.bbc.com/news/articles/cgerwp7rdlvo},
    urldate = {2025-08-27},
    author = {Yousif, Nadine},
    month = aug,
    year = {2025},
}

@article{heinz_randomized_2025,
    title = {Randomized {Trial} of a {Generative} {AI} {Chatbot} for {Mental} {Health} {Treatment}},
    volume = {2},
    url = {https://ai.nejm.org/doi/full/10.1056/AIoa2400802},
    doi = {10.1056/AIoa2400802},
    number = {4},
    urldate = {2025-06-27},
    journal = {NEJM AI},
    author = {Heinz, Michael V. and Mackin, Daniel M. and Trudeau, Brianna M. and Bhattacharya, Sukanya and Wang, Yinzhou and Banta, Haley A. and Jewett, Abi D. and Salzhauer, Abigail J. and Griffin, Tess Z. and Jacobson, Nicholas C.},
    month = mar,
    year = {2025},
    note = {Publisher: Massachusetts Medical Society},
    pages = {AIoa2400802},
}

@article{stade_large_2024,
    title = {Large language models could change the future of behavioral healthcare: a proposal for responsible development and evaluation},
    volume = {3},
    copyright = {2024 The Author(s)},
    issn = {2731-4251},
    shorttitle = {Large language models could change the future of behavioral healthcare},
    url = {https://www.nature.com/articles/s44184-024-00056-z},
    doi = {10.1038/s44184-024-00056-z},
    language = {en},
    number = {1},
    urldate = {2024-06-19},
    journal = {npj Mental Health Research},
    author = {Stade, Elizabeth C. and Stirman, Shannon Wiltsey and Ungar, Lyle H. and Boland, Cody L. and Schwartz, H. Andrew and Yaden, David B. and Sedoc, João and DeRubeis, Robert J. and Willer, Robb and Eichstaedt, Johannes C.},
    month = apr,
    year = {2024},
    note = {Publisher: Nature Publishing Group},
    pages = {1--12},
}

@article{ma_integrating_2024,
    title = {Integrating large language models in mental health practice: a qualitative descriptive study based on expert interviews},
    volume = {12},
    issn = {2296-2565},
    shorttitle = {Integrating large language models in mental health practice},
    url = {https://www.frontiersin.org/journals/public-health/articles/10.3389/fpubh.2024.1475867/full},
    doi = {10.3389/fpubh.2024.1475867},
    language = {English},
    urldate = {2025-06-20},
    journal = {Frontiers in Public Health},
    author = {Ma, Yingzhuo and Zeng, Yi and Liu, Tong and Sun, Ruoshan and Xiao, Mingzhao and Wang, Jun},
    month = nov,
    year = {2024},
}

@article{lawrence_opportunities_2024,
    title = {The {Opportunities} and {Risks} of {Large} {Language} {Models} in {Mental} {Health}},
    volume = {11},
    copyright = {Unless stated otherwise, all articles are open-access distributed under the terms of the Creative Commons Attribution License (http://creativecommons.org/licenses/by/2.0/), which permits unrestricted use, distribution, and reproduction in any medium, provided the original work ("first published in the Journal of Medical Internet Research...") is properly cited with original URL and bibliographic citation information. The complete bibliographic information, a link to the original publication on http://www.jmir.org/, as well as this copyright and license information must be included.},
    url = {https://mental.jmir.org/2024/1/e59479},
    doi = {10.2196/59479},
    language = {EN},
    number = {1},
    urldate = {2024-10-21},
    journal = {JMIR Mental Health},
    author = {Lawrence, Hannah R. and Schneider, Renee A. and Rubin, Susan B. and Matarić, Maja J. and McDuff, Daniel J. and Bell, Megan Jones},
    month = jul,
    year = {2024},
    note = {Company: JMIR Mental Health
Distributor: JMIR Mental Health
Institution: JMIR Mental Health
Label: JMIR Mental Health
Publisher: JMIR Publications Inc., Toronto, Canada},
    pages = {e59479},
}

@article{hatch_when_2025,
    title = {When {ELIZA} meets therapists: {A} {Turing} test for the heart and mind},
    volume = {2},
    issn = {2837-8156},
    shorttitle = {When {ELIZA} meets therapists},
    url = {https://journals.plos.org/mentalhealth/article?id=10.1371/journal.pmen.0000145},
    doi = {10.1371/journal.pmen.0000145},
    language = {en},
    number = {2},
    urldate = {2025-02-17},
    journal = {PLOS Mental Health},
    author = {Hatch, S. Gabe and Goodman, Zachary T. and Vowels, Laura and Hatch, H. Dorian and Brown, Alyssa L. and Guttman, Shayna and Le, Yunying and Bailey, Benjamin and Bailey, Russell J. and Esplin, Charlotte R. and Harris, Steven M. and Jr, D. Payton Holt and McLaughlin, Merranda and O’Connell, Patrick and Rothman, Karen and Ritchie, Lane and Jr, D. Nicholas Top and Braithwaite, Scott R.},
    month = feb,
    year = {2025},
    note = {Publisher: Public Library of Science},
    pages = {e0000145},
}

@article{martin_conversational_2023,
    title = {Conversational agents for {Children}'s mental health and mental disorders: {A} scoping review},
    volume = {1},
    issn = {2949-8821},
    shorttitle = {Conversational agents for {Children}'s mental health and mental disorders},
    url = {https://www.sciencedirect.com/science/article/pii/S2949882123000282},
    doi = {10.1016/j.chbah.2023.100028},
    number = {2},
    urldate = {2025-02-18},
    journal = {Computers in Human Behavior: Artificial Humans},
    author = {Martin, Rachael and Richmond, Sally},
    month = aug,
    year = {2023},
    pages = {100028},
}

@misc{openai_building_2025,
    title = {Building more helpful {ChatGPT} experiences for everyone},
    url = {https://openai.com/index/building-more-helpful-chatgpt-experiences-for-everyone/},
    language = {en-US},
    urldate = {2025-09-27},
    author = {{OpenAI}},
    month = sep,
    year = {2025},
}

@article{mahmoud_identifying_2025,
    title = {Identifying the {Public}'s {Beliefs} {About} {Generative} {Artificial} {Intelligence}: {A} {Big} {Data} {Approach}},
    volume = {72},
    issn = {1558-0040},
    shorttitle = {Identifying the {Public}'s {Beliefs} {About} {Generative} {Artificial} {Intelligence}},
    url = {https://ieeexplore.ieee.org/document/10854904},
    doi = {10.1109/TEM.2025.3534088},
    urldate = {2025-06-05},
    journal = {IEEE Transactions on Engineering Management},
    author = {Mahmoud, Ali B. and Kumar, V and Spyropoulou, Stavroula},
    year = {2025},
    pages = {827--841},
}

@book{bender_ai_2025,
    title = {The {AI} {Con}: {How} to {Fight} {Big} {Tech}’s {Hype} and {Create} the {Future} {We} {Want}},
    isbn = {978-1-5299-4989-6},
    shorttitle = {The {AI} {Con}},
    language = {en},
    publisher = {Random House},
    author = {Bender, Emily M. and Hanna, Alex},
    month = may,
    year = {2025},
    note = {Google-Books-ID: T3EXEQAAQBAJ},
}

@article{herbener_are_2025,
    title = {Are lonely youngsters turning to chatbots for companionship? {The} relationship between chatbot usage and social connectedness in {Danish} high-school students},
    volume = {196},
    issn = {1071-5819},
    shorttitle = {Are lonely youngsters turning to chatbots for companionship?},
    url = {https://www.sciencedirect.com/science/article/pii/S1071581924001927},
    doi = {10.1016/j.ijhcs.2024.103409},
    urldate = {2025-01-23},
    journal = {International Journal of Human-Computer Studies},
    author = {Herbener, Arthur Bran and Damholdt, Malene Flensborg},
    month = feb,
    year = {2025},
    pages = {103409},
}

@article{coeckelbergh_can_2012,
    title = {Can we trust robots?},
    volume = {14},
    issn = {1572-8439},
    url = {https://doi.org/10.1007/s10676-011-9279-1},
    doi = {10.1007/s10676-011-9279-1},
    language = {en},
    number = {1},
    urldate = {2024-03-23},
    journal = {Ethics and Information Technology},
    author = {Coeckelbergh, Mark},
    month = mar,
    year = {2012},
    pages = {53--60},
}

@article{kreps_exploring_2023,
    title = {Exploring the artificial intelligence “{Trust} paradox”: {Evidence} from a survey experiment in the {United} {States}},
    volume = {18},
    issn = {1932-6203},
    shorttitle = {Exploring the artificial intelligence “{Trust} paradox”},
    url = {https://journals.plos.org/plosone/article?id=10.1371/journal.pone.0288109},
    doi = {10.1371/journal.pone.0288109},
    language = {en},
    number = {7},
    urldate = {2025-02-03},
    journal = {PLOS ONE},
    author = {Kreps, Sarah and George, Julie and Lushenko, Paul and Rao, Adi},
    month = jul,
    year = {2023},
    note = {Publisher: Public Library of Science},
    pages = {e0288109},
}

@article{minina_jeunemaitre_ai_2025,
    title = {{AI} lovers, friends and partners: consumer imagination work in {AI} humanization},
    volume = {0},
    issn = {1025-3866},
    shorttitle = {{AI} lovers, friends and partners},
    url = {https://doi.org/10.1080/10253866.2025.2505013},
    doi = {10.1080/10253866.2025.2505013},
    number = {0},
    urldate = {2025-10-03},
    journal = {Consumption Markets \& Culture},
    author = {Minina Jeunemaître, Alisa and Masè, Stefania and Smith, Jamie},
    month = may,
    year = {2025},
    note = {Publisher: Routledge
\_eprint: https://doi.org/10.1080/10253866.2025.2505013},
    pages = {1--21},
}

@article{kirk_benefits_2024,
    title = {The benefits, risks and bounds of personalizing the alignment of large language models to individuals},
    volume = {6},
    copyright = {2024 Springer Nature Limited},
    issn = {2522-5839},
    url = {https://www.nature.com/articles/s42256-024-00820-y},
    doi = {10.1038/s42256-024-00820-y},
    language = {en},
    number = {4},
    urldate = {2025-10-03},
    journal = {Nature Machine Intelligence},
    author = {Kirk, Hannah Rose and Vidgen, Bertie and Röttger, Paul and Hale, Scott A.},
    month = apr,
    year = {2024},
    note = {Publisher: Nature Publishing Group},
    pages = {383--392},
}

@misc{kalai_why_2025,
    title = {Why {Language} {Models} {Hallucinate}},
    url = {http://arxiv.org/abs/2509.04664},
    doi = {10.48550/arXiv.2509.04664},
    urldate = {2025-09-13},
    author = {Kalai, Adam Tauman and Nachum, Ofir and Vempala, Santosh S. and Zhang, Edwin},
    month = sep,
    year = {2025},
    note = {arXiv:2509.04664 [cs]},
}

@inproceedings{zheng_customizing_2025,
    address = {New York, NY, USA},
    series = {{CHI} '25},
    title = {Customizing {Emotional} {Support}: {How} {Do} {Individuals} {Construct} and {Interact} {With} {LLM}-{Powered} {Chatbots}},
    isbn = {979-8-4007-1394-1},
    shorttitle = {Customizing {Emotional} {Support}},
    url = {https://dl.acm.org/doi/10.1145/3706598.3713453},
    doi = {10.1145/3706598.3713453},
    urldate = {2025-10-01},
    booktitle = {Proceedings of the 2025 {CHI} {Conference} on {Human} {Factors} in {Computing} {Systems}},
    publisher = {Association for Computing Machinery},
    author = {Zheng, Xi and Li, Zhuoyang and Gui, Xinning and Luo, Yuhan},
    month = apr,
    year = {2025},
    pages = {1--20},
}

@book{de_waal_age_2010,
    title = {The {Age} of {Empathy}: {Nature}'s {Lessons} for a {Kinder} {Society}},
    isbn = {978-0-7710-2738-3},
    shorttitle = {The {Age} of {Empathy}},
    language = {en},
    publisher = {McClelland \& Stewart},
    author = {de Waal, Frans},
    month = sep,
    year = {2010},
    note = {Google-Books-ID: 2\_qOEAAAQBAJ},
}

@article{langford_social_1997,
    title = {Social support: a conceptual analysis},
    volume = {25},
    issn = {1365-2648},
    shorttitle = {Social support},
    url = {https://onlinelibrary.wiley.com/doi/abs/10.1046/j.1365-2648.1997.1997025095.x},
    doi = {10.1046/j.1365-2648.1997.1997025095.x},
    language = {en},
    number = {1},
    urldate = {2025-01-24},
    journal = {Journal of Advanced Nursing},
    author = {Langford, Catherine Penny Hinson and Bowsher, Juanita and Maloney, Joseph P. and Lillis, Patricia P.},
    year = {1997},
    note = {\_eprint: https://onlinelibrary.wiley.com/doi/pdf/10.1046/j.1365-2648.1997.1997025095.x},
    pages = {95--100},
}

@book{frank_persuasion_2025,
    title = {Persuasion and {Healing}: {A} {Comparative} {Study} of {Psychotherapy}},
    isbn = {978-1-4214-5138-1},
    shorttitle = {Persuasion and {Healing}},
    language = {en},
    publisher = {JHU Press},
    author = {Frank, Jerome D. and Frank, Julia B. and Wampold, Bruce E.},
    month = mar,
    year = {2025},
    note = {Google-Books-ID: so4UEQAAQBAJ},
}

@inproceedings{pendse_treatment_2022,
    address = {New York, NY, USA},
    series = {{CHI} '22},
    title = {From {Treatment} to {Healing}: {Envisioning} a {Decolonial} {Digital} {Mental} {Health}},
    isbn = {978-1-4503-9157-3},
    shorttitle = {From {Treatment} to {Healing}},
    url = {https://dl.acm.org/doi/10.1145/3491102.3501982},
    doi = {10.1145/3491102.3501982},
    urldate = {2024-10-21},
    booktitle = {{CHI} {Conference} on {Human} {Factors} in {Computing} {Systems}},
    publisher = {Association for Computing Machinery},
    author = {Pendse, Sachin R and Nkemelu, Daniel and Bidwell, Nicola J and Jadhav, Sushrut and Pathare, Soumitra and De Choudhury, Munmun and Kumar, Neha},
    month = apr,
    year = {2022},
    pages = {1--23},
}

@article{wainberg_challenges_2017,
    title = {Challenges and {Opportunities} in {Global} {Mental} {Health}: a {Research}-to-{Practice} {Perspective}},
    volume = {19},
    issn = {1535-1645},
    shorttitle = {Challenges and {Opportunities} in {Global} {Mental} {Health}},
    url = {https://doi.org/10.1007/s11920-017-0780-z},
    doi = {10.1007/s11920-017-0780-z},
    language = {en},
    number = {5},
    urldate = {2025-10-04},
    journal = {Current Psychiatry Reports},
    author = {Wainberg, Milton L. and Scorza, Pamela and Shultz, James M. and Helpman, Liat and Mootz, Jennifer J. and Johnson, Karen A. and Neria, Yuval and Bradford, Jean-Marie E. and Oquendo, Maria A. and Arbuckle, Melissa R.},
    month = apr,
    year = {2017},
    pages = {28},
}

@article{henderson_mental_2013,
    title = {Mental {Illness} {Stigma}, {Help} {Seeking}, and {Public} {Health} {Programs}},
    volume = {103},
    issn = {0090-0036},
    url = {https://ajph.aphapublications.org/doi/full/10.2105/AJPH.2012.301056},
    doi = {10.2105/AJPH.2012.301056},
    number = {5},
    urldate = {2025-10-04},
    journal = {American Journal of Public Health},
    author = {Henderson, Claire and Evans-Lacko, Sara and Thornicroft, Graham},
    month = may,
    year = {2013},
    note = {Publisher: American Public Health Association},
    pages = {777--780},
}

@book{office_of_the_surgeon_general_osg_our_2023,
    address = {Washington (DC)},
    series = {Publications and {Reports} of the {Surgeon} {General}},
    title = {Our {Epidemic} of {Loneliness} and {Isolation}: {The} {U}.{S}. {Surgeon} {General}’s {Advisory} on the {Healing} {Effects} of {Social} {Connection} and {Community}},
    shorttitle = {Our {Epidemic} of {Loneliness} and {Isolation}},
    url = {http://www.ncbi.nlm.nih.gov/books/NBK595227/},
    language = {eng},
    urldate = {2025-10-04},
    publisher = {US Department of Health and Human Services},
    author = {{Office of the Surgeon General}},
    year = {2023},
    pmid = {37792968},
}

@article{surkalim_prevalence_2022,
    title = {The prevalence of loneliness across 113 countries: systematic review and meta-analysis},
    volume = {376},
    copyright = {© Author(s) (or their employer(s)) 2019. Re-use permitted under CC BY-NC. No commercial re-use. See rights and permissions. Published by BMJ.. http://creativecommons.org/licenses/by-nc/4.0/This is an Open Access article distributed in accordance with the Creative Commons Attribution Non Commercial (CC BY-NC 4.0) license, which permits others to distribute, remix, adapt, build upon this work non-commercially, and license their derivative works on different terms, provided the original work is properly cited and the use is non-commercial. See: http://creativecommons.org/licenses/by-nc/4.0/.},
    issn = {1756-1833},
    shorttitle = {The prevalence of loneliness across 113 countries},
    url = {https://www.bmj.com/content/376/bmj-2021-067068},
    doi = {10.1136/bmj-2021-067068},
    language = {en},
    urldate = {2025-10-04},
    journal = {BMJ},
    author = {Surkalim, Daniel L. and Luo, Mengyun and Eres, Robert and Gebel, Klaus and Buskirk, Joseph van and Bauman, Adrian and Ding, Ding},
    month = feb,
    year = {2022},
    pmid = {35140066},
    note = {Publisher: British Medical Journal Publishing Group
Section: Research},
    pages = {e067068},
}

@article{olsson_teens_2025,
    title = {Teens are increasingly turning to {AI} companions, and it could be harming them},
    url = {http://theconversation.com/teens-are-increasingly-turning-to-ai-companions-and-it-could-be-harming-them-261955},
    urldate = {2025-08-23},
    author = {Olsson, Craig and Spry, Liz},
    month = aug,
    year = {2025},
    doi = {10.64628/AA.seteyqwd5},
}

@article{ng_deepseek_2025,
    title = {'{DeepSeek} brought me to tears': {How} young {Chinese} find therapy in {AI}},
    shorttitle = {'{DeepSeek} brought me to tears'},
    url = {https://www.bbc.com/news/articles/cy7g45g2nxno},
    urldate = {2025-02-13},
    journal = {BBC},
    author = {Ng, Kelly},
    month = feb,
    year = {2025},
}

@article{pirnay_we_2023,
    title = {We {Spoke} to {People} {Who} {Started} {Using} {ChatGPT} {As} {Their} {Therapist}},
    url = {https://www.vice.com/en/article/z3mnve/we-spoke-to-people-who-started-using-chatgpt-as-their-therapist},
    urldate = {2024-04-25},
    journal = {Vice},
    author = {Pirnay, Emma},
    month = apr,
    year = {2023},
}

@article{tong_digital_2022,
    title = {Digital {Therapeutic} {Alliance} {With} {Fully} {Automated} {Mental} {Health} {Smartphone} {Apps}: {A} {Narrative} {Review}},
    volume = {13},
    issn = {1664-0640},
    shorttitle = {Digital {Therapeutic} {Alliance} {With} {Fully} {Automated} {Mental} {Health} {Smartphone} {Apps}},
    url = {https://www.frontiersin.org/journals/psychiatry/articles/10.3389/fpsyt.2022.819623/full},
    doi = {10.3389/fpsyt.2022.819623},
    language = {English},
    urldate = {2024-10-21},
    journal = {Frontiers in Psychiatry},
    author = {Tong, Fangziyun and Lederman, Reeva and D'Alfonso, Simon and Berry, Katherine and Bucci, Sandra},
    month = jun,
    year = {2022},
    note = {Publisher: Frontiers},
}

@article{torous_focusing_2023,
    title = {Focusing on {Digital} {Research} {Priorities} for {Advancing} the {Access} and {Quality} of {Mental} {Health}},
    volume = {10},
    issn = {2368-7959},
    doi = {10.2196/47898},
    language = {eng},
    journal = {JMIR mental health},
    author = {Torous, John and Benson, Nicole M. and Myrick, Keris and Eysenbach, Gunther},
    month = apr,
    year = {2023},
    pmid = {37093624},
    pmcid = {PMC10167575},
    pages = {e47898},
}

@article{goldberg_mobile_2022,
    title = {Mobile phone-based interventions for mental health: {A} systematic meta-review of 14 meta-analyses of randomized controlled trials},
    volume = {1},
    issn = {2767-3170},
    shorttitle = {Mobile phone-based interventions for mental health},
    doi = {10.1371/journal.pdig.0000002},
    language = {eng},
    number = {1},
    journal = {PLOS digital health},
    author = {Goldberg, Simon B. and Lam, Sin U. and Simonsson, Otto and Torous, John and Sun, Shufang},
    year = {2022},
    pmid = {35224559},
    pmcid = {PMC8881800},
    pages = {e0000002},
}

@article{linardon_current_2024,
    title = {Current evidence on the efficacy of mental health smartphone apps for symptoms of depression and anxiety. {A} meta-analysis of 176 randomized controlled trials},
    volume = {23},
    copyright = {© 2024 World Psychiatric Association.},
    issn = {2051-5545},
    url = {https://onlinelibrary.wiley.com/doi/abs/10.1002/wps.21183},
    doi = {10.1002/wps.21183},
    language = {en},
    number = {1},
    urldate = {2025-10-04},
    journal = {World Psychiatry},
    author = {Linardon, Jake and Torous, John and Firth, Joseph and Cuijpers, Pim and Messer, Mariel and Fuller-Tyszkiewicz, Matthew},
    year = {2024},
    note = {\_eprint: https://onlinelibrary.wiley.com/doi/pdf/10.1002/wps.21183},
    pages = {139--149},
}

@article{haque_overview_2023,
    title = {An {Overview} of {Chatbot}-{Based} {Mobile} {Mental} {Health} {Apps}: {Insights} {From} {App} {Description} and {User} {Reviews}},
    volume = {11},
    issn = {2291-5222},
    shorttitle = {An {Overview} of {Chatbot}-{Based} {Mobile} {Mental} {Health} {Apps}},
    url = {https://www.ncbi.nlm.nih.gov/pmc/articles/PMC10242473/},
    doi = {10.2196/44838},
    urldate = {2024-10-07},
    journal = {JMIR mHealth and uHealth},
    author = {Haque, M D Romael and Rubya, Sabirat},
    month = may,
    year = {2023},
    pmid = {37213181},
    pmcid = {PMC10242473},
    pages = {e44838},
}

@article{lee_artificial_2025,
    title = {Artificial intelligence conversational agents in mental health: {Patients} see potential, but prefer humans in the loop},
    volume = {15},
    issn = {1664-0640},
    shorttitle = {Artificial intelligence conversational agents in mental health},
    url = {https://www.frontiersin.org/journals/psychiatry/articles/10.3389/fpsyt.2024.1505024/full},
    doi = {10.3389/fpsyt.2024.1505024},
    language = {English},
    urldate = {2025-10-04},
    journal = {Frontiers in Psychiatry},
    author = {Lee, Hyein S. and Wright, Colton and Ferranto, Julia and Buttimer, Jessica and Palmer, Clare E. and Welchman, Andrew and Mazor, Kathleen M. and Fisher, Kimberly A. and Smelson, David and O’Connor, Laurel and Fahey, Nisha and Soni, Apurv},
    month = jan,
    year = {2025},
    note = {Publisher: Frontiers},
}

@article{guo_large_2024,
    title = {Large {Language} {Model} for {Mental} {Health}: {A} {Systematic} {Review}},
    shorttitle = {Large {Language} {Model} for {Mental} {Health}},
    url = {https://mental.jmir.org/2024/1/e57400},
    doi = {https://doi.org/10.2196/57400},
    urldate = {2024-06-03},
    journal = {JMIR Mental Health},
    author = {Guo, Zhijun and Lai, Alvina and Thygesen, Johan Hilge and Farrington, Joseph and Keen, Thomas and Li, Kezhi},
    month = oct,
    year = {2024},
}

@article{brittain_google_2025,
    chapter = {Boards, Policy \& Regulation},
    title = {Google, {AI} firm must face lawsuit filed by a mother over suicide of son, {US} court says},
    url = {https://www.reuters.com/sustainability/boards-policy-regulation/google-ai-firm-must-face-lawsuit-filed-by-mother-over-suicide-son-us-court-says-2025-05-21/},
    language = {en},
    urldate = {2025-06-16},
    journal = {Reuters},
    author = {Brittain, Blake},
    month = may,
    year = {2025},
}

@misc{papiernik_safety_2025,
    title = {Safety {Evaluation} of a {Clinical}-{Grade} {Generative} {AI} {Agent} for {Anxiety} and {Depression} {Symptoms}},
    url = {https://osf.io/8kvm6_v1},
    doi = {10.31234/osf.io/8kvm6_v1},
    language = {en-us},
    urldate = {2025-10-02},
    author = {Papiernik, Pearla and Dietz, Alyssa and McMahon, Molly and Buttimer, Jessica and Ewbank, Michael and Lowrie, Ria and Ambrose, Pete and Marshall, Emily and Droog-Hayes, Max and Beasley, Joanna and Mehew, Shaun and Catarino, Ana and Cummins, Ronan and Tablan, Valentin and Blackwell, Andrew and Palmer, Clare},
    month = sep,
    year = {2025},
}

@article{benk_twenty-four_2024,
    title = {Twenty-four years of empirical research on trust in {AI}: a bibliometric review of trends, overlooked issues, and future directions},
    issn = {1435-5655},
    shorttitle = {Twenty-four years of empirical research on trust in {AI}},
    url = {https://doi.org/10.1007/s00146-024-02059-y},
    doi = {10.1007/s00146-024-02059-y},
    language = {en},
    urldate = {2025-01-31},
    journal = {AI \& SOCIETY},
    author = {Benk, Michaela and Kerstan, Sophie and von Wangenheim, Florian and Ferrario, Andrea},
    month = oct,
    year = {2024},
}

@article{vereschak_how_2021,
    title = {How to {Evaluate} {Trust} in {AI}-{Assisted} {Decision} {Making}? {A} {Survey} of {Empirical} {Methodologies}},
    volume = {5},
    shorttitle = {How to {Evaluate} {Trust} in {AI}-{Assisted} {Decision} {Making}?},
    url = {https://dl.acm.org/doi/10.1145/3476068},
    doi = {10.1145/3476068},
    number = {CSCW2},
    urldate = {2024-11-06},
    journal = {Proceedings of the ACM on Human-Computer Interaction},
    author = {Vereschak, Oleksandra and Bailly, Gilles and Caramiaux, Baptiste},
    month = oct,
    year = {2021},
    pages = {327:1--327:39},
}

@article{bach_systematic_2024,
    title = {A {Systematic} {Literature} {Review} of {User} {Trust} in {AI}-{Enabled} {Systems}: {An} {HCI} {Perspective}},
    volume = {40},
    issn = {1044-7318},
    shorttitle = {A {Systematic} {Literature} {Review} of {User} {Trust} in {AI}-{Enabled} {Systems}},
    url = {https://doi.org/10.1080/10447318.2022.2138826},
    doi = {10.1080/10447318.2022.2138826},
    number = {5},
    urldate = {2024-11-06},
    journal = {International Journal of Human–Computer Interaction},
    author = {Bach, Tita Alissa and Khan, Amna and Hallock, Harry and Beltrão, Gabriela and Sousa, Sonia},
    month = mar,
    year = {2024},
    note = {Publisher: Taylor \& Francis
\_eprint: https://doi.org/10.1080/10447318.2022.2138826},
    pages = {1251--1266},
}

@article{frauenberger_entanglement_2019,
    title = {Entanglement {HCI} {The} {Next} {Wave}?},
    volume = {27},
    issn = {1073-0516},
    url = {https://dl.acm.org/doi/10.1145/3364998},
    doi = {10.1145/3364998},
    number = {1},
    urldate = {2025-02-07},
    journal = {ACM Transactions on Computer-Human Interaction},
    author = {Frauenberger, Christopher},
    month = nov,
    year = {2019},
    pages = {2:1--2:27},
}

@article{zhan_healthcare_2024,
    title = {Healthcare {Voice} {AI} {Assistants}: {Factors} {Influencing} {Trust} and {Intention} to {Use}},
    volume = {8},
    shorttitle = {Healthcare {Voice} {AI} {Assistants}},
    url = {https://dl.acm.org/doi/10.1145/3637339},
    doi = {10.1145/3637339},
    number = {CSCW1},
    urldate = {2024-11-02},
    journal = {Proceedings of the ACM on Human-Computer Interaction},
    author = {Zhan, Xiao and Abdi, Noura and Seymour, William and Such, Jose},
    month = apr,
    year = {2024},
    pages = {62:1--62:37},
}

@article{glikson_human_2020,
    title = {Human {Trust} in {Artificial} {Intelligence}: {Review} of {Empirical} {Research}},
    volume = {14},
    issn = {1941-6520},
    shorttitle = {Human {Trust} in {Artificial} {Intelligence}},
    url = {https://journals.aom.org/doi/10.5465/annals.2018.0057},
    doi = {10.5465/annals.2018.0057},
    number = {2},
    urldate = {2024-01-24},
    journal = {Academy of Management Annals},
    author = {Glikson, Ella and Woolley, Anita Williams},
    month = jul,
    year = {2020},
    note = {Publisher: Academy of Management},
    pages = {627--660},
}

@inproceedings{becker_factors_2024,
    address = {Berlin, Heidelberg},
    title = {Factors of {Trust} {Building} in {Conversational} {AI} {Systems}: {A} {Literature} {Review}},
    isbn = {978-3-031-60613-7},
    shorttitle = {Factors of {Trust} {Building} in {Conversational} {AI} {Systems}},
    url = {https://doi.org/10.1007/978-3-031-60611-3_3},
    doi = {10.1007/978-3-031-60611-3_3},
    urldate = {2024-07-01},
    booktitle = {International {Conference} on {Human}-{Computer} {Interaction}},
    publisher = {Springer-Verlag},
    author = {Becker, Cornelia and Fischer, Mahsa},
    month = jun,
    year = {2024},
    pages = {27--44},
}

@article{jeon_effects_2024,
    title = {The {Effects} of {Emotions} on {Trust} in {Human}-{Computer} {Interaction}: {A} {Survey} and {Prospect}},
    volume = {40},
    issn = {1044-7318},
    shorttitle = {The {Effects} of {Emotions} on {Trust} in {Human}-{Computer} {Interaction}},
    url = {https://doi.org/10.1080/10447318.2023.2261727},
    doi = {10.1080/10447318.2023.2261727},
    number = {22},
    urldate = {2025-06-05},
    journal = {International Journal of Human–Computer Interaction},
    author = {Jeon, Myounghoon},
    month = nov,
    year = {2024},
    note = {Publisher: Taylor \& Francis
\_eprint: https://doi.org/10.1080/10447318.2023.2261727},
    pages = {6864--6882},
}

@article{huynh_generative_2025,
    title = {In generative artificial intelligence we trust: unpacking determinants and outcomes for cognitive trust},
    issn = {1435-5655},
    shorttitle = {In generative artificial intelligence we trust},
    url = {https://doi.org/10.1007/s00146-025-02378-8},
    doi = {10.1007/s00146-025-02378-8},
    language = {en},
    urldate = {2025-10-05},
    journal = {AI \& SOCIETY},
    author = {Huynh, Minh-Tay and Aichner, Thomas},
    month = may,
    year = {2025},
}

@inproceedings{wang_understanding_2025,
    address = {New York, NY, USA},
    series = {{CHI} '25},
    title = {Understanding {Attitudes} and {Trust} of {Generative} {AI} {Chatbots} for {Social} {Anxiety} {Support}},
    isbn = {979-8-4007-1394-1},
    url = {https://dl.acm.org/doi/10.1145/3706598.3714286},
    doi = {10.1145/3706598.3714286},
    urldate = {2025-10-05},
    booktitle = {Proceedings of the 2025 {CHI} {Conference} on {Human} {Factors} in {Computing} {Systems}},
    publisher = {Association for Computing Machinery},
    author = {Wang, Yimeng and Wang, Yinzhou and Crace, Kelly and Zhang, Yixuan},
    month = apr,
    year = {2025},
    pages = {1--21},
}

@article{bordin_generalizability_1979,
    title = {The generalizability of the psychoanalytic concept of the working alliance},
    volume = {16},
    issn = {0033-3204},
    doi = {10.1037/h0085885},
    number = {3},
    journal = {Psychotherapy: Theory, Research \& Practice},
    author = {Bordin, Edward S.},
    year = {1979},
    note = {Place: US
Publisher: Division of Psychotherapy (29), American Psychological Association},
    pages = {252--260},
}

@misc{liu_trustworthy_2024,
    title = {Trustworthy {LLMs}: a {Survey} and {Guideline} for {Evaluating} {Large} {Language} {Models}' {Alignment}},
    shorttitle = {Trustworthy {LLMs}},
    url = {http://arxiv.org/abs/2308.05374},
    doi = {10.48550/arXiv.2308.05374},
    urldate = {2024-06-04},
    publisher = {arXiv},
    author = {Liu, Yang and Yao, Yuanshun and Ton, Jean-Francois and Zhang, Xiaoying and Guo, Ruocheng and Cheng, Hao and Klochkov, Yegor and Taufiq, Muhammad Faaiz and Li, Hang},
    month = mar,
    year = {2024},
    note = {arXiv:2308.05374 [cs]},
}

@article{tan_ai_2025,
    title = {{AI} meets psychology: an exploratory study of large language models’ competence in psychotherapy contexts},
    volume = {1},
    issn = {null},
    shorttitle = {{AI} meets psychology},
    url = {https://doi.org/10.1080/29974100.2025.2545258},
    doi = {10.1080/29974100.2025.2545258},
    number = {1},
    urldate = {2025-10-08},
    journal = {Journal of Psychology and AI},
    author = {Tan, Kean Sian and Cervin, Matti and Leman, Patrick and Nielsen, Kristopher and Kumar, Prashanth Vasantha and Medvedev, Oleg},
    month = dec,
    year = {2025},
    note = {Publisher: Taylor \& Francis
\_eprint: https://doi.org/10.1080/29974100.2025.2545258},
    pages = {2545258},
}

@article{ma_understanding_2024,
    title = {Understanding the {Benefits} and {Challenges} of {Using} {Large} {Language} {Model}-based {Conversational} {Agents} for {Mental} {Well}-being {Support}},
    volume = {2023},
    issn = {1942-597X},
    url = {https://www.ncbi.nlm.nih.gov/pmc/articles/PMC10785945/},
    urldate = {2024-06-28},
    journal = {AMIA Annual Symposium Proceedings},
    author = {Ma, Zilin and Mei, Yiyang and Su, Zhaoyuan},
    month = jan,
    year = {2024},
    pmid = {38222348},
    pmcid = {PMC10785945},
    pages = {1105--1114},
}

@article{knight_openai_nodate,
    title = {{OpenAI} {Scrambles} to {Update} {GPT}-5 {After} {Users} {Revolt}},
    issn = {1059-1028},
    url = {https://www.wired.com/story/openai-gpt-5-backlash-sam-altman/},
    language = {en-US},
    urldate = {2025-10-10},
    journal = {Wired},
    author = {Knight, Will},
    note = {Section: tags},
    year = {2025},
}

@incollection{luhmann_familiarity_1988,
    title = {Familiarity, {Confidence}, {Trust}: {Problems} and {Alternatives}},
    shorttitle = {Familiarity, {Confidence}, {Trust}},
    url = {https://www.semanticscholar.org/paper/Familiarity%2C-Confidence%2C-Trust%3A-Problems-and-Luhmann/a5ae78f779284090b9cadb9c5b05501c223f9c23},
    urldate = {2024-04-04},
    booktitle = {Trust: making and breaking cooperative relations},
    author = {Luhmann, Niklas},
    year = {1988},
    pages = {94--107},
}

@article{ayad_attributions_2025,
    title = {Attributions of intent and moral responsibility to {AI} agents},
    volume = {3},
    issn = {2949-8821},
    url = {https://www.sciencedirect.com/science/article/pii/S2949882124000677},
    doi = {10.1016/j.chbah.2024.100107},
    urldate = {2025-02-27},
    journal = {Computers in Human Behavior: Artificial Humans},
    author = {Ayad, Reem and Plaks, Jason E.},
    month = mar,
    year = {2025},
    pages = {100107},
}

@article{yin_ai_2024,
    title = {{AI} can help people feel heard, but an {AI} label diminishes this impact},
    volume = {121},
    url = {https://www.pnas.org/doi/abs/10.1073/pnas.2319112121},
    doi = {10.1073/pnas.2319112121},
    number = {14},
    urldate = {2025-02-13},
    journal = {Proceedings of the National Academy of Sciences},
    author = {Yin, Yidan and Jia, Nan and Wakslak, Cheryl J.},
    month = apr,
    year = {2024},
    pages = {e2319112121},
}

@article{glickman_how_2025,
    title = {How human–{AI} feedback loops alter human perceptual, emotional and social judgements},
    volume = {9},
    copyright = {2024 The Author(s)},
    issn = {2397-3374},
    url = {https://www.nature.com/articles/s41562-024-02077-2},
    doi = {10.1038/s41562-024-02077-2},
    language = {en},
    number = {2},
    urldate = {2025-08-23},
    journal = {Nature Human Behaviour},
    author = {Glickman, Moshe and Sharot, Tali},
    month = feb,
    year = {2025},
    pages = {345--359},
}

@article{vowels_are_2024,
    title = {Are chatbots the new relationship experts? {Insights} from three studies},
    volume = {2},
    issn = {2949-8821},
    shorttitle = {Are chatbots the new relationship experts?},
    url = {https://www.sciencedirect.com/science/article/pii/S2949882124000379},
    doi = {10.1016/j.chbah.2024.100077},
    number = {2},
    urldate = {2025-02-18},
    journal = {Computers in Human Behavior: Artificial Humans},
    author = {Vowels, Laura M.},
    month = aug,
    year = {2024},
    pages = {100077},
}

@article{hoffman_understanding_2024,
    title = {Understanding young adults’ attitudes towards using {AI} chatbots for psychotherapy: {The} role of self-stigma},
    volume = {2},
    issn = {2949-8821},
    shorttitle = {Understanding young adults’ attitudes towards using {AI} chatbots for psychotherapy},
    url = {https://www.sciencedirect.com/science/article/pii/S294988212400046X},
    doi = {10.1016/j.chbah.2024.100086},
    number = {2},
    urldate = {2025-02-18},
    journal = {Computers in Human Behavior: Artificial Humans},
    author = {Hoffman, Benjamin David and Oppert, Michelle Leanne and Owen, Mikaela},
    month = aug,
    year = {2024},
    pages = {100086},
}

@article{araujo_speaking_2024,
    title = {From speaking like a person to being personal: {The} effects of personalized, regular interactions with conversational agents},
    volume = {2},
    issn = {2949-8821},
    shorttitle = {From speaking like a person to being personal},
    url = {https://www.sciencedirect.com/science/article/pii/S2949882123000300},
    doi = {10.1016/j.chbah.2023.100030},
    number = {1},
    urldate = {2025-02-18},
    journal = {Computers in Human Behavior: Artificial Humans},
    author = {Araujo, Theo and Bol, Nadine},
    month = jan,
    year = {2024},
    pages = {100030},
}

@article{brunswicker_trust_2025,
    title = {Trust through words: {The} systemize-empathize-effect of language in task-oriented conversational agents},
    volume = {165},
    issn = {0747-5632},
    shorttitle = {Trust through words},
    url = {https://www.sciencedirect.com/science/article/pii/S0747563224003844},
    doi = {10.1016/j.chb.2024.108516},
    urldate = {2025-02-05},
    journal = {Computers in Human Behavior},
    author = {Brunswicker, Sabine and Zhang, Yifan and Rashidian, Christopher and Linna, Daniel W.},
    month = apr,
    year = {2025},
    pages = {108516},
}

@misc{freitas_emotional_2025,
    title = {Emotional {Manipulation} by {AI} {Companions}},
    url = {http://arxiv.org/abs/2508.19258},
    doi = {10.48550/arXiv.2508.19258},
    urldate = {2025-08-30},
    author = {Freitas, Julian De and Oğuz-Uğuralp, Zeliha and Kaan-Uğuralp, Ahmet},
    month = aug,
    year = {2025},
    note = {arXiv:2508.19258 [cs]},
}

@misc{weidinger_sociotechnical_2023,
    title = {Sociotechnical {Safety} {Evaluation} of {Generative} {AI} {Systems}},
    url = {http://arxiv.org/abs/2310.11986},
    doi = {10.48550/arXiv.2310.11986},
    urldate = {2024-09-13},
    author = {Weidinger, Laura and Rauh, Maribeth and Marchal, Nahema and Manzini, Arianna and Hendricks, Lisa Anne and Mateos-Garcia, Juan and Bergman, Stevie and Kay, Jackie and Griffin, Conor and Bariach, Ben and Gabriel, Iason and Rieser, Verena and Isaac, William},
    month = oct,
    year = {2023},
    note = {arXiv:2310.11986 [cs]},
}

@article{mckelvey_stable_2024,
    title = {({Un})stable diffusions: {The} publics, publicities, and publicizations of generative {AI}},
    volume = {6},
    copyright = {Copyright (c) 2024 Fenwick McKelvey, Joanna Redden, Jonathan Roberge, Luke Stark},
    issn = {2003-1998},
    shorttitle = {({Un})stable diffusions},
    url = {https://publicera.kb.se/jdsr/article/view/40453},
    doi = {10.33621/jdsr.v6i440453},
    language = {en},
    number = {4},
    urldate = {2025-10-10},
    journal = {Journal of Digital Social Research},
    author = {McKelvey, Fenwick and Redden, Joanna and Roberge, Jonathan and Stark, Luke},
    month = dec,
    year = {2024},
    pages = {1--12},
}

@article{fluckiger_alliance_2018,
    title = {The alliance in adult psychotherapy: {A} meta-analytic synthesis.},
    volume = {55},
    copyright = {http://www.apa.org/pubs/journals/resources/open-access.aspx},
    issn = {1939-1536, 0033-3204},
    shorttitle = {The alliance in adult psychotherapy},
    url = {https://doi.apa.org/doi/10.1037/pst0000172},
    doi = {10.1037/pst0000172},
    language = {en},
    number = {4},
    urldate = {2024-10-18},
    journal = {Psychotherapy},
    author = {Flückiger, Christoph and Del Re, A. C. and Wampold, Bruce E. and Horvath, Adam O.},
    month = dec,
    year = {2018},
    pages = {316--340},
}

@inproceedings{selbst_fairness_2019,
    address = {New York, NY, USA},
    series = {{FAT}* '19},
    title = {Fairness and {Abstraction} in {Sociotechnical} {Systems}},
    isbn = {978-1-4503-6125-5},
    url = {https://dl.acm.org/doi/10.1145/3287560.3287598},
    doi = {10.1145/3287560.3287598},
    urldate = {2025-01-28},
    booktitle = {{ACM} {Conference} on {Fairness}, {Accountability}, and {Transparency}},
    publisher = {Association for Computing Machinery},
    author = {Selbst, Andrew D. and Boyd, Danah and Friedler, Sorelle A. and Venkatasubramanian, Suresh and Vertesi, Janet},
    month = jan,
    year = {2019},
    pages = {59--68},
}

@misc{kleinberg_textwash_2022,
    title = {Textwash -- automated open-source text anonymisation},
    url = {http://arxiv.org/abs/2208.13081},
    doi = {10.48550/arXiv.2208.13081},
    urldate = {2025-05-21},
    publisher = {arXiv},
    author = {Kleinberg, Bennett and Davies, Toby and Mozes, Maximilian},
    month = aug,
    year = {2022},
    note = {arXiv:2208.13081 [cs]},
}

@techreport{robb_talk_2025,
    address = {San Francisco},
    title = {Talk, {Trust}, and {Trade}-{Offs}: {How} and {Why} {Teens} {Use} {AI} {Companions} {\textbar} {Common} {Sense} {Media}},
    shorttitle = {Talk, {Trust}, and {Trade}-{Offs}},
    url = {https://www.commonsensemedia.org/research/talk-trust-and-trade-offs-how-and-why-teens-use-ai-companions},
    language = {en},
    urldate = {2025-11-12},
    institution = {Common Sense Media},
    author = {Robb, Michael B. and {Supreet Mann}},
    month = jul,
    year = {2025},
}

@article{petzel_prejudiced_2025,
    title = {Prejudiced interactions with large language models ({LLMs}) reduce trustworthiness and behavioral intentions among members of stigmatized groups},
    volume = {165},
    issn = {0747-5632},
    url = {https://www.sciencedirect.com/science/article/pii/S074756322500010X},
    doi = {10.1016/j.chb.2025.108563},
    urldate = {2025-02-05},
    journal = {Computers in Human Behavior},
    author = {Petzel, Zachary W. and Sowerby, Leanne},
    month = apr,
    year = {2025},
    pages = {108563},
}

\appendix
\section{Participants Demographics} \label{participants-demographics}
\begin{table}[H]
  \caption{Demographic information of the 24 participants that completed the study.}
  \label{tab:participants-demographics}
  \begin{tabular}{llll}
    \toprule
    \textit{Question}&Value&N&\%\\
    \midrule
    \textit{Age} & 18-24&  12&50\% \\
        & 25-34&  9&38\% \\
        & 35-44&  3&13\% \\
    \arrayrulecolor{lightgray}
    \hline
    \textit{Gender}  & Woman& 12&50\% \\
            & Man& 9&38\% \\
            & Astralgender& 1&4\% \\
            & Non-binary& 1&4\% \\
            & Prefer not to say& 1&4\% \\
    \hline
    \textit{Country} & UK& 18&75\% \\
            & US& 2&8\% \\
            & India& 2&8\% \\
            & China& 1&4\% \\
            & Italy& 1&4\% \\
    \hline
    \textit{Location} & Urban& 17&71\% \\
            & Suburban& 7&29\% \\
    \hline
    \textit{Ethnicity}   & Black & 8&33\% \\
                & Indian& 5&21\% \\
                & White& 4&17\% \\
                & Chinese& 2&8\% \\
                & Asian& 1&4\% \\
                & Arab& 1&4\% \\
                & African& 1&4\% \\
                & British & 1&4\% \\
                & Prefer not to say& 1&4\% \\
    \hline
    \textit{Education}   & Bachelor&  14&58\% \\
                & Master&  5&21\% \\
                & Secondary school&  3&13\% \\
                & College& 2&8\% \\
    \hline
    \textit{Employment}  & Student& 10&42\% \\
                & Part-time& 9&38\% \\
                & Full-time& 2&8\% \\
                & Self-employed& 1&4\% \\
                & Not currently employed& 1&4\% \\
                & Prefer not to say& 1&4\% \\
    \arrayrulecolor{black}
    \bottomrule
\end{tabular}
\end{table}
\section{Study Materials}
\subsection{Diary Study} \label{diary-questions}
\begin{enumerate}
    \item Which generative AI did you chat with?
    \item Please copy/paste the content of your chat (optional)
    \item What did you chat about?
    \item What led you to seek support specifically from a generative AI for this situation?
    \item Briefly describe the context of this chat
    \item Briefly describe what you experienced (feelings, thoughts, behaviours)
    \item What aspects of the AI responses influenced your trust?
    \item What aspects of the wider AI system were present in your mind during this chat?
    \item What made you end the chat and how did you feel afterwards? 
    \item Overall, how satisfied were you with the support received? (5-point scale)
    \item Overall, how much did you trust the chatbot during this chat? (5-point scale)
    \item During this chat, did you use special features to provide more context to the AI?
    \item If you provided any custom instructions, please briefly describe them:
\end{enumerate}

\subsection{Interview Guide} \label{interview-guide}
\begin{enumerate}
  \item The Meaning of Trust and AI:
  \begin{itemize}
    \item Using your own words, could you describe what Generative AI is?
    \item What does trusting Generative AI mean to you?
    \item What does trusting Generative AI for emotional support mean to you?
  \end{itemize}

  \item Personal Experiences:
  \begin{itemize}
    \item Can you tell me more about your experiences of emotional support from Generative AI?
    \item How has your use changed over time since you started?
    \item How have you customised the AIs you use?
    \item What makes you decide to start and end a conversation?
    \item How do you feel during these conversations?
  \end{itemize}

  \item Behaviours of Generative AI:
  \begin{itemize}
    \item How do the AI responses generally compare with your expectations?
    \item Has there been a recent time where AI surprised you negatively or positively?
    \item When AI responds to your request, how do you know that you can trust that information?
  \end{itemize}

  \item Context Beyond Immediate Interactions:
  \begin{itemize}
    \item How do you like your environment to be when you chat with AI?
    \item How do you feel about the opinions of your friends and family regarding seeking emotional support from generative AI?
    \item How do you feel about how the media portrays seeking emotional support from it?
  \end{itemize}

  \item Social and Material Aspects of the AI Ecosystem:
  \begin{itemize}
    \item When you ask something to the AI or share your experiences or feelings, what do you think shapes how it responds to you?
    \item How would you assign responsibility when generative AI gives responses that may cause harm?
  \end{itemize}

  \item Conclusion:
  \begin{itemize}
    \item What else is important to you about the topic we have been talking about?
    \item What else would you like to share about the topic we have been talking about?
  \end{itemize}
\end{enumerate}

\end{document}